\documentclass[aps,prb,preprint,showpacs,showkeys]{revtex4-1}

\usepackage[pdftex]{graphicx}
\usepackage[caption=false]{subfig}
\usepackage{color,soul}

\begin{document}
\title{\emph{ab initio} Electronic Transport Model with Explicit Solution to the ​Linearized Boltzmann Transport Equation}%
\author{Alireza Faghaninia}%
\affiliation{Department of Energy, Environmental, and Chemical Engineering, Washington University, St. Louis, Missouri 63130, USA}
\author{Joel W. Ager III}
\affiliation{Materials Sciences Division, Lawrence Berkeley National Laboratory, Berkeley, California 94720, USA}
\author{Cynthia S. Lo}%
\email[]{clo@wustl.edu}
\affiliation{Department of Energy, Environmental, and Chemical Engineering, Washington University, St. Louis, Missouri 63130, USA }
\date{\today}%

\begin{abstract}
Accurate models of carrier transport are essential for describing the electronic properties of semiconductor materials. To the best of our knowledge, the current models following the framework of the Boltzmann transport equation (BTE) either rely heavily on experimental data (i.e., semi-empirical), or utilize simplifying assumptions, such as the constant relaxation time approximation (BTE-cRTA). While these models offer valuable physical insights and accurate calculations of transport properties in some cases, they often lack sufficient accuracy -- particularly in capturing the correct trends with temperature and carrier concentration. We present here a transport model for calculating low-field electrical drift mobility and Seebeck coefficient of n-type semiconductors, by explicitly considering relevant physical phenomena (i.e. elastic and inelastic scattering mechanisms). We first rewrite expressions for the rates of elastic scattering mechanisms, in terms of \emph{ab initio} properties, such as the band structure, density of states, and polar optical phonon frequency.  We then solve the linear BTE to obtain the perturbation to the electron distribution -- resulting from the dominant scattering mechanisms -- and use this to calculate the overall mobility and Seebeck coefficient. Using our model, we accurately calculate electrical transport properties of the compound n-type semiconductors, GaAs and InN, over various ranges of temperature and carrier concentration.  Our fully predictive model provides high accuracy when compared to experimental measurements on both GaAs and InN, and vastly outperforms both semi-empirical models and the BTE-cRTA.  Therefore, we assert that this approach represents a first step towards a fully \emph{ab initio} carrier transport model that is valid in all compound semiconductors.
\end{abstract}



\maketitle

\section{Introduction}
\label{introduction}

Accurate models of carrier transport are essential for describing the electronic properties of semiconductor materials, which are particularly important for clean energy applications ranging from photovoltaics to thermoelectrics to photoelectrocatalysts. There has been an increased focus on using compound semiconductors, including those that are degenerately and heavily doped, for these applications.  To better understand existing materials and discover new ones, a fully predictive model that correlates electronic structure to properties is essential.  Unfortunately, to the best of our knowledge, no model, based on \emph{ab initio} calculations, currently exists to fully capture the elastic and inelastic scattering effects of charge carriers; as a result, errors arise when utilizing the current models. While an \emph{ab initio} model will certainly improve our understanding of the carrier transport mechanism(s) in existing semiconductors, it can also aid in the search for high-performing materials by improving the accuracy of high-throughput computations \cite{carrier24,carrier25}.

There currently exist two main categories of models, based on the Boltzmann transport equation (BTE), for calculating the conductivity and Seebeck coefficient of semiconductors that are governed by band conduction. The first category of BTE-based models are commonly known as single parabolic band models, even though the treatment of the conduction band may not be explicitly parabolic. These models can be described as "semi-empirical", since experimentally measured parameters, such as the electron or hole effective mass, band gap, dielectric constant and polar optical (PO) phonon frequency, are used in closed-form expressions for the various scattering rates. \textcolor{black}{Note that the overall mobility due to elastic scattering is calculated by averaging, according to Matthiessen's rule, the mobilities due to each scattering contribution. The main adjustable parameter in these models is the effective mass, which can be varied to fit the calculated transport properties to the experimental measurements.} While such models often impressively capture the changes in properties over various ranges of temperature and carrier concentration, they are restricted to the materials for which experimental data are available; therefore, the predictability of such models are very limited. 

There are numerous examples of models in this category \cite{book8,carrier32,carrier1,carrier26,carrier18}, such as that by Ehrenreich \cite{carrier26}, who modeled the GaAs band structure and PO-phonon scattering by reviewing the experimental data \cite{carrier26}, and that by Sankey et al. \cite{carrier1}, who considered the effects of resonance, ionized impurity, and polar optical phonon scattering in GaAs.  In these models, all of the scattering mechanisms are commonly treated \textcolor{black}{using the relaxation time approximation (RTA); here, the relaxation time is written as a power law function of energy -- thus, the details of elastic and inelastic scattering (e.g., PO phonon) captured by the \emph{ab initio} band structure are disregarded. Scattering rates, particularly inelastic ones, have already been shown to not follow such power law distributions \cite{book8,Ager_InN}, so the basic assumptions fail. Even in cases where the BTE is explicitly solved for PO phonon and the perturbation to the electronic distribution is obtained without the RTA assumption}, the results are still heavily dependent on available experimental data. As an example, Miller et al. \cite{Ager_InN} used the latter approach to calculate the mobility and Seebeck coefficient of InN samples, which had been grown by molecular beam epitaxy (MBE) and plasma assisted MBE so that all exhibited heteroepitaxial growth with linear charged dislocations; thus, these dislocations were found to be the limiting scattering mechanism.   

The second category of BTE-based models relies on the \emph{ab initio} band structure of the material, rather than specific experimentally measured parameters, but generally utilizes the relaxation time approximation to the BTE (BTE-RTA) as a simplification. Restrepo et al. \cite{carrier84} calculated the mobility of n-doped silicon at different electron concentrations in BTE-RTA and \emph{ab initio} framework where electron-phonon interactions are treated as elastic with the electron distribution unchanged from the equilibrium Fermi-Dirac. On the other hand, the constant relaxation time approximation (BTE-cRTA) simplifies the equation even more, which enables closed form expressions for both conductivity divided by relaxation time and Seebeck coefficient. The advantage of these models is the ability to calculate properties of new materials, for which experimental data is unavailable. This type of model works well for some materials for which the relaxation time is fairly constant, as evidenced by the work of Madsen and Singh \cite{50}. However, inelastic scattering mechanisms \textcolor{black}{change the electron energy and directly affect the distribution. Lumping all the elastic and inelastic scattering mechanisms into a single constant and assuming an equilibrium Fermi-Dirac distribution in BTE-cRTA framework greatly damages the predictive ability of such models; as an example, transport properties in some cases are very far from experimental measurements}. Furthermore, the relaxation time constant is usually determined by fitting the calculated conductivity to experimental data. \textcolor{black}{It should be noted that the calculation of this constant is not necessary when calculating the Seebeck coefficient. This is due to the simplifying assumptions that the relaxation time is both a constant and direction independent \cite{50} which does not always hold.}
Therefore, BTE-cRTA suffers not only from inaccuracy in predicting the changes of properties with temperature or carrier concentration in many materials, but also from lack of pure predictability since it still relies on experimental data for the computation of the relaxation time.

Instead, we propose that accurate calculations of electronic transport properties, within the Boltzmann transport framework, are possible by combining relevant treatment of the elastic and inelastic scattering mechanisms with \emph{ab initio} calculations of the electronic and phonon band structures. Ultimately, an \emph{ab initio} theory for carrier transport will need to qualitatively and quantitatively predict trends in material properties, such as conductivity and Seebeck coefficient, as a function of temperature or carrier concentration.  Validation of the theory against experimentally measured properties will thus give insight into which scattering effects are dominant. 

In this paper, we present a band transport model for calculating low-field electrical drift mobility and Seebeck coefficient of n-type semiconductors.  We then validate our model by calculating the properties of two III-V semiconductors, GaAs and InN, with different carrier concentrations over various temperatures, and comparing them to experimental values as well as those calculated using the other transport models described above.  We choose these materials because the \emph{ab initio} band structure of GaAs is similar to those used in the earlier semi-empirical models \textcolor{black}{at it can be reasonably well described with a single band model}, whereas the \emph{ab initio} band structure of InN \textcolor{black}{and the limiting scattering mechanisms are} quite different; thus, these two materials allow us to bracket the range of expected behavior of our proposed model.

\section{Carrier Transport Model}
\label{theory}

\subsection{Solution to the Boltzmann Transport Equation}

In order to calculate the mobility and Seebeck coefficient, we solve the Boltzmann transport equation (BTE) using Rode's iterative method  \cite{book8,Ager_InN,rode_physrevb_1970,carrier19,fundamental_of_carrier_transport_2009,semiconductor_transport_2000,carrier60,carrier63,carrier75} (Appendix \ref{appendix:btesolution}) to obtain the electron distribution in response to a small driving force (e.g. a small electric field or a small temperature gradient). It is important to note that we do not use the relaxation time approximation (RTA) in solving the BTE, so neither a variable nor a constant relaxation time appears in this expression.  Due to the assumption of a small driving force, we aim to calculate only the linear response to the perturbation; thus, the general form of the electron distribution remains the \textcolor{black}{at equilibrium} Fermi-Dirac distribution. We can then write:
\begin{equation}
f\left(\mathbf{k}\right)=f_0\left[\varepsilon \left(k\right)\right]+xg\left(k\right)
\label{eq:perturbation}
\end{equation}
where $f$ is the actual distribution of the electrons, including both elastic and inelastic scattering mechanisms, $f_0$ is the equilibrium Fermi-Dirac distribution, $x$ is the cosine of the angle between the small driving force and $\mathbf{k}$, $g\left(k\right)$ is the perturbation to the distribution caused by the small driving force and finally $k=|\mathbf{k}|$. For the sake of simplicity, we express the conduction band as the average energy of the electrons as a function of distance, $k$, from the conduction band minimum (CBM) which is often at the center of the Brillouin Zone (i.e. $\Gamma$ point); furthermore, we assume that the small driving force is aligned with $k$ (i.e., $x$=1). Although this is similar in spirit to the isotropic band assumption, we take the anisotropy into account by averaging the energy values of the \emph{ab initio} calculated band structure, $\varepsilon \left(\mathbf{k}\right)$, as a function of $k$ rather than explicitly including $\mathbf{k}$ in every direction. Alternatively, if we wish to consider the directional transport properties, we can include the calculated band structure only in that specific direction. Here, we will focus on calculating and reporting the overall average mobility and Seebeck coefficient.

Our goal is to calculate the perturbation to the distribution \cite{book8}, $g\left(k\right)$. In \textcolor{black}{the reformulated Boltzmann transport equation shown in Equation \ref{eq:g}}, there are scattering-in, $S_i \left( g \right)$, and scattering-out, $S_o$, \textcolor{black}{terms for inelastic scattering mechanisms}. However, these terms also depend, in turn, on the electronic distribution as well as elastic scattering rates, $\nu_{el}$. Therefore, the \textcolor{black}{BTE} must be solved self-consistently to obtain $g\left(k\right)$:
\begin{equation}
g\left(k\right) = \frac{S_i \left[ g(k) \right] - v\left(k\right) \left( \frac{\partial f}{\partial z} \right) - \frac{eE}{\hbar} \ \left( \frac{\partial f}{\partial k} \right)}{S_o(k) + \nu_{el}(k)}
\label{eq:g}
\end{equation}
where $E$ is the low electric field \textcolor{black}{and $v\left(k\right)$ is the electron group velocity. The derivation of the BTE in the form shown in Equation \ref{eq:g} can be found in the literature\cite{book8}}. The inelastic scattering mechanism that tends to dominate at room temperature is polar optical (PO) phonon scattering, for which we have provided the description of the $S_i \left( g \right)$ and $S_o$ terms in Equations \ref{eq:bte_reduced_inelastic_in} and \ref{eq:bte_reduced_inelastic_out}. The influence of inelastic scattering mechanisms on $g$, and therefore the overall mobility, are captured through the terms $S_i \left( g \right)$ and $S_o$ in Equation \ref{eq:g}, while elastic scattering mechanisms affect the overall mobility by the term $\nu_{el}$. This term is the sum of all elastic scattering rates inside the material, it can be evaluated according to Matthiessen's rule: 
\begin{equation} 
\nu_{el}\left(k\right) =\nu_{ii}\left(k\right)+\nu_{pe}\left(k\right)+\nu_{de}\left(k\right)+\nu_{dis}\left(k\right)
\end{equation}
where the subscripts $el$, $ii$, $pe$, $de$, and $dis$ stand for elastic, ionized impurity, piezoelectric, deformation potential and dislocation scattering rates, respectively. \textcolor{black}{Therefore, the effect of relevant elastic and inelastic scattering mechanisms are taken into account by explicitly solving the BTE (Equation \ref{eq:g}) to obtain $g\left(k\right)$}. 

When calculating various properties, several terms in Equation \ref{eq:g} will be set to zero. For a Seebeck coefficient, $S$, calculation, the applied electric driving force, $- \left( \frac{eE}{\hbar} \right) \left( \frac{\partial f}{\partial \mathbf{k}} \right)$, is set to zero.  Only the thermal driving force, $v \left( \frac{\partial f}{\partial z} \right)$, in Equation \ref{eq:g} is taken into consideration when calculating the perturbation to the electron distribution \cite{book8}. Assuming a uniform electron concentration over the space at which a small temperature difference exists, the Seebeck coefficient is \cite{book8}:
\begin{equation}
S = \frac{k_B}{e}\left[\frac{\varepsilon_F}{k_BT}-\frac{\int k^2f \left( 1-f \right) \left( \frac{\varepsilon}{k_BT} \right) dk}{\int k^2f \left( 1-f \right) dk} \right]-\frac{\frac{J}{\sigma}}{\frac{\partial T}{\partial z}}
\label{thermopower}
\end{equation}  

For a mobility calculation, the applied thermal driving force in Equation \ref{eq:g} is set to zero, so that only the contribution of the electric driving force is included. The mobility is:
\begin{equation}
\label{eq:mukp}
\mu = \frac{1}{3}\frac{\int v \left( k \right) \left( \frac{k}{\pi} \right)^2 \left( \frac{g}{E} \right) dk}{\int \left( \frac{k}{\pi} \right)^2 f dk}
\end{equation}
Note that in Equation \ref{eq:mukp}, the free electron density of states, $\left(\frac{k}{\pi}\right)^2$, has been used, which would limit its applicability in compound semiconductors.  Thus, the replacement of this term by its \emph{ab initio}-calculated counterpart would greatly improve the accuracy of the resulting mobility. Furthermore, the scalar group velocity, $v \left( k \right)$, is used since the energy is averaged as a function of distance from the $\Gamma$ point. In general, we use the band structure, density of state, electron group velocity, conduction band wavefunction admixture and PO phonon frequency in calculating the mobility and Seebeck coefficient. Therefore, \textcolor{black}{all} of the required inputs to Equation \ref{eq:mukp} are calculated \emph{ab initio}, which greatly enhances the predictability of the model. In other words, the main difference between our proposed carrier transport model and previous semi-empirical models \cite{book8,book8_2,carrier32,carrier1,carrier26,Ager_InN,rode_physrevb_1970,carrier19,fundamental_of_carrier_transport_2009,semiconductor_transport_2000,carrier61,carrier62} is the use of \emph{ab initio} parameters instead of experimentally measured electron effective mass, band gap, etc., which eliminates the need for theories such as $\mathbf{k}\cdot\mathbf{p}$ to describe the nonparabolicity or anisotropy of the conduction band. Instead, for calculating the overlap integral, we express the conduction band wavefunction as a linear combination of s-type and p-type basis functions, with coefficients of $a$ and $c$, respectively \cite{book8}. \textcolor{black}{These coefficients can be directly calculated \emph{ab initio} without the need to assume an s-like conduction band wavefunction (i.e., no assumption of a parabolic band).}

The rates of the elastic scattering mechanisms, $\nu_{el}$, are calculated from the electron group velocities, $v$, and density of states, $D_S$; thus, the mobility may be calculated directly from the electronic \textcolor{black}{band} structure.  The original form of these equations from $\mathbf{k}\cdot\mathbf{p}$ theory, and the modified equations that we propose, are listed in Table \ref{tab:elastic_scattering_expressions}. Note that in every equation, $\frac{\hbar k}{m d\left(k\right)}$, which, in semi-empirical models, is the group velocity fitted to experiment by the band gap and effective mass of the semiconductor (included in $d\left(k\right)$, see Table \ref{tab:elastic_scattering_expressions}), has been replaced by its \emph{ab initio} counterpart, or $v\left(k\right)$, which is calculated directly from the band structure.

\begin{table}
	\caption{\label{tab:elastic_scattering_expressions}The original equations \cite{book8,Ager_InN}, based on $\mathbf{k}\cdot\mathbf{p}$ theory for elastic scattering rates and overall drift mobility, and proposed modifications, based on \emph{ab initio} parameters, introduced in this work.}
	\begin{ruledtabular}
	\begin{tabular}{cc}
		 $\mathbf{k} \cdot \mathbf{p}$ theory with empirical parameters & \emph{ab initio} \\ 
		 \hline 
		 \footnotemark[1] $\nu_{ii}\left(k\right)=\frac{e^4Nmd \left( k \right)}{8\pi\epsilon_0^2\hbar^3k^3}\left[D\left(k\right) \ln\left(1+\frac{4k^2}{\beta^2}\right)-B \left(k\right)\right]$ & $\nu_{ii}\left(k\right)=\frac{e^4N}{8\pi\epsilon_0^2\hbar^2k^2v\left(k\right)}\left[D\left(k\right) \ln\left(1+\frac{4k^2}{\beta^2}\right)-B\left(k\right)\right]$ \\
		 $\beta^2=\frac{e^2}{\epsilon_0 k_BT}\int \left(\frac{k}{\pi}\right)^2f\left(1-f\right)dk$  & $\beta^2=\frac{e^2}{\epsilon_0 k_BT}\int D_S\left(\varepsilon\right)f\left(1-f\right)d\varepsilon$
		\\  
		 $\nu_{pe}\left(k\right)=\frac{e^2k_BTP^2md\left(k\right)}{6\pi\hbar^3\epsilon_0k} \left[ 3-6c^2\left(k\right)+4c^4\left(k\right) \right]$& $\nu_{pe}\left(k\right)=\frac{e^2k_BTP^2}{6\pi\hbar^2\epsilon_0v\left(k\right)}(3-6c^2\left(k\right)+4c^4\left(k\right))$\\
		\footnotemark[2] $c^2\left(k\right)=1-\frac{1+\alpha}{2\alpha}$, $\alpha^2\left(k\right)=1+\frac{2\hbar^2k^2}{m\varepsilon_g}\left(\frac{m}{m^\ast}-1\right)$ & $c\left(k\right)$ : obtained directly from wavefunctions 
		\\ 
		$\nu_{de}\left(k\right)=\frac{e^2k_BTE_D^2mkd\left(k\right)}{3\pi\hbar^3c_{el}}\left[3-8c^2\left(k\right)+6c^4\left(k\right)\right]$& $\nu_{de}\left(k\right)=\frac{e^2k_BTE_D^2k^2}{3\pi\hbar^2c_{el}v\left(k\right)}\left[3-8c^2\left(k\right)+6c^4\left(k\right)\right]$
		\\  
		 $\nu_{dis}\left(k\right)=\frac{N_{dis}e^4md\left(k\right)}{\hbar^3\epsilon_0^2c_l^2}\frac{1}{\left(1+\frac{4k^2}{\beta^2}\right)^{3/2}\beta^4}$ , $\frac{1}{d\left(k\right)}= 1+\frac{m/m^\ast-1}{\alpha}$& $\nu_{dis}\left(k\right)=\frac{N_{dis}e^4k}{\hbar^2\epsilon_0^2c_l^2v\left(k\right)}\frac{1}{\left(1+\frac{4k^2}{\beta^2}\right)^{3/2}\beta^4}$ \\
		 \hline
		 $\mu_{overall}=\frac{\hbar}{3m}\frac{\int k^3(g\left(k\right)/Ed\left(k\right))dk}{\int k^2fdk}$ & $\mu_{overall}=\frac{1}{3E}\frac{\int v(\varepsilon)D_S\left(\varepsilon\right)g\left(\varepsilon\right)d\varepsilon}{\int D_S\left(\varepsilon\right)f\left(\varepsilon\right)d\varepsilon}$  \\
		 $g\left(k\right)=f\left(k\right)-f_0\left(k\right)$ &  $g\left(\varepsilon\right)=f\left(\varepsilon\right)-f_0\left(\varepsilon\right)$ \\
	\end{tabular} 
	\end{ruledtabular}
	\footnotetext[1]{The subscripts stand for: $ii$ (ionized impurity), $pe$ (piezoelectric acoustic phonon), $de$ (deformation), and $dis$ (charged dislocation scattering). The parameters are: $m$ (electron mass), $m^\ast$ (effective mass), $\epsilon_0$ (low-frequency dielectric constant), $\varepsilon_g$ (band gap), $v\left(k\right)$ (electron group velocity), $D_S\left(\varepsilon\right)$ (\emph{ab initio} calculated density of states), $c\left(k\right)$ (contribution of p-type orbitals to the conduction band), $\beta$ (inverse ionized impurity charge screening length), $E_D$ (deformation potential), $c_l$ (lattice constant), $E$ (small electric field), and $c_{el}$ (spherically averaged elastic constant). $B\left(k\right)$ and $D\left(k\right)$ are just collection of the parameters: $c$, $k$ and $\beta$. Their purpose is to simplify the equation\cite{book8}.}
	\footnotetext[2]{The $c(k)$ parameter is the contribution of the $p$ orbital to the wavefunction of the band.  In the $\mathbf{k} \cdot \mathbf{p}$ formulation, it has a closed-form expression that includes the band gap and experimental effective mass.  In the \emph{ab initio} formulation, this wavefunction admixture can be calculated by projecting the wavefunctions onto spherical harmonics that are nonzero within the sphere around each ion; this procedure is already implemented in the Vienna \emph{ab initio} Simulation Package (VASP) \cite{vasp1,vasp2,vasp3,vasp4}.}
\end{table}

As an example, the DFT-calculated density of states (DOS) can be plugged into Equation \ref{eq:iiscatteringbeta} to obtain the inverse charge screening length, $\beta$, in ionized impurity scattering. Furthermore, the numerator and denominator of the integrand in Equation \ref{eq:mukp} both contain the density of states of a free electron gas, $\left( \frac{k}{\pi} \right)^2$.  Since this can also be calculated \emph{ab initio} for the specific system of interest, $D_S$ can instead be substituted in the equation for calculating the mobility and reformulated in terms of the energies, $\varepsilon$:
\begin{equation}
\mu = \frac{1}{3E}\frac{\int v\left( \varepsilon \right) D_S \left( \varepsilon \right) g \left( \varepsilon \right) d \varepsilon}{\int D_S \left( \varepsilon \right) f \left( \varepsilon \right) d \varepsilon}
\label{eq:mobility_Ds}
\end{equation}
where, again, $v\left(k\right)$ is the electron group velocity and $g$ is the perturbation to the electron distribution, which is calculated iteratively using Equation \ref{eq:g}, and can be expressed both as a function of $k$ or $\varepsilon\left(k\right)$ (i.e., the band structure).

Once the mobilities of the electrons and holes are known, the electrical conductivity can be readily calculated:
\begin{equation}
\sigma = ne\mu_e + pe\mu_h
\label{eq:conductivity_simple}
\end{equation}
where $n$ and $p$ are the concentration of electrons and holes, respectively, $e$ is the absolute value of the charge of an electron and $\mu_e$ and $\mu_h$ are the mobility of electrons and holes respectively. 

It should be noted that there are fundamental differences between the model that we have presented here and those relying on the relaxation time approximation (RTA), and particularly, BTE-cRTA. Rather than simplification of the collision term in the BTE (Equation \ref{eq:btesmallE}) through the RTA (Equation \ref{eq:perturbation}), we fully involve this term by considering both elastic and inelastic scattering mechanisms. \textcolor{black}{It is noteworthy that the BTE-cRTA formulation only implicitly takes into account elastic and inelastic scattering mechanisms, by fitting the overall relaxation time to experimental data with no explicit consideration of changes in electron distribution from each type of scattering mechanism}. Furthermore, unlike the semi-empirical models that were described above, we use \emph{ab initio} parameters; thus, higher predictability and little to no dependence on experimental data is achieved.

\subsection{\emph{ab initio} Parameters}

The main input that is needed for the transport model is the crystal structure of the semiconductor material, from which \emph{ab initio} parameters, such as the (optimized) lattice constant, PO phonon frequency, dielectric \textcolor{black}{and piezoelectric constants}, deformation potential and effective mass, can be computed. 

We also need to know the Fermi level to compute scattering rates in Table \ref{tab:elastic_scattering_expressions}. In order to obtain the Fermi level, we calculate the carrier concentration and match it to the given concentration (input), $n$, according to Equation \ref{eq:e-concentration}:
\begin{equation}
n=\frac{1}{V} \int_{\varepsilon_c}^{+\infty} g \left(\varepsilon \right)f\left(\varepsilon \right)d\varepsilon
\label{eq:e-concentration}
\end{equation}
Since both of the III-V semiconductors considered here are n-type, the concentration of hole carriers is negligible.  The concentration of ionized impurities, $N_{ii}$ (see Table \ref{tab:elastic_scattering_expressions}), is the sum of the concentration of all ionized centers regardless of the sign of their charge, since they are scatterer centers in both cases \cite{carrier33}:
\begin{equation}
N_{ii} = N_A + N_D + \frac{N_{dis}}{c_l}
\label{eq:N_ii}
\end{equation}
where $N_D$ and $N_A$ are concentration of donors and acceptors, respectively. $N_{ii}$ can then be calculated at a given electron concentration, $n$, by iteratively solving the charge balance equation \cite{Ager_InN}:
\begin{equation}
n + N_A = N_D + \frac{N_{dis}}{c_l}
\label{eq:charge_balance}
\end{equation}

where the density of dislocations, $N_{dis}$, is only relevant for InN and is considered to be zero for GaAs. In both GaAs and InN, temperatures lower than 20 K need not be considered due to the deionization of shallow donors at lower temperatures, as observed experimentally \cite{carrier20}. In the case of InN, electronic scattering from existing linear charged dislocations thus becomes important. The density of the dislocations, $N_{dis}$, can be determined from TEM images, in the units of $\left(\textrm{cm}^{-2}\right)$. We can thus obtain the overall density in bulk, by assuming that these linear dislocations are uniformly developed along the $c$-axis. This is reflected in dividing the dislocation density by the lattice constant, $c_l$, in Equation \ref{eq:charge_balance}. By doing that, we are assuming that there is one unit of positive charge (donor) per unit cell. For InN samples, according to Miller et al. \cite{Ager_InN}, one can assume full ionization of the donors, and therefore, a compensation level of one (i.e., $\frac{N_D+N_A}{n}=1$ or $N_D>>N_A$). Also, the assumption of donor or acceptor charged dislocations yields similar results \cite{Ager_InN}; therefore, we assume donor dislocations dominate here. It should be noted that we compare the calculated $N_{dis}$ with the corresponding experimental data if available; otherwise, the limit for electronic properties at different values of $N_{dis}$ can be calculated without the need for experimental data. 

On the other hand, in a pure, epitaxially-grown, high-mobility GaAs sample with an electron concentration of $n=3.0 \times 10^{13}$, no dislocations exist (i.e. $N_{dis}=0$). The concentrations of donors and acceptors have been separately reported \cite{carrier20,book8}, so this provides validation of the accuracy of our model, without needing to solve for $N_{ii}$.  However, in the general case where the electron concentration is unknown, we can plot the mobility and Seebeck coefficient at different compensation ratios to define the limit of the transport properties, as shown in Figure \ref{fig:GaAs-s}. Therefore, it is important to note that only when comparing with experimental mobilities/Seebeck coefficients do we use experimentally measured electron concentrations; otherwise, we may calculate \emph{ab initio} mobility or Seebeck coefficient, for example, at various electron concentrations, without any reliance on experimental data (e.g., as shown in Figure \ref{fig:GaAs-s}). 

We use Brooks-Herring theory for singly-charged ionized impurity scattering \cite{carrier33}, as shown in Table \ref{tab:elastic_scattering_expressions}. This is supported by the fact that in GaAs, oxygen impurities, O$_{As}^{+1}$, have been confirmed to be dominant and singly charged \cite{carrier34}, while in InN, nitrogen (donor) vacancies, V$_N^{+1}$, are dominant and singly charged \cite{carrier74}. It should be noted that the Brooks-Herring formulation is more accurate at low carrier concentrations, since at high concentrations, despite the inherent assumption of the theory, not all electrons are screened by the charge of an ionized center. More information on the Brooks-Herring ionized impurity model is available in Appendix \ref{appendix:btesolution}.

In order to calculate the low- and high-frequency dielectric constants, we use density functional perturbation theory (DFPT), as implemented in VASP, to determine Born effective charges, dielectric and piezoelectric tensors, including local field effects in DFT, as well as the force-constant matrices and internal strain tensors. We then subtract the ionic contribution to the static dielectric tensor to obtain the high-frequency dielectric constant \cite{carrier72,carrier73}. Furthermore, the inelastic scattering effect is strongly dependent on the longitudinal polar optical phonon frequencies, $\omega_{po}$.  These frequencies can be calculated using the Phonopy code \cite{phonopy}, where we identify the highest energy peak in the optical phonon density of states. It should be noted that at and around the $\Gamma$ point, the phonon frequency is almost constant (Figure \ref{fig:phonon}). 

To calculate \emph{ab initio} the deformation potential, $E_D$, we strain the system and calculate the energy of the conduction band of InN and GaAs unit cells at different volumes. Then, we approximate the deformation potential using the following equation:
\begin{equation}
E_D=\left.-V\left( \frac{\partial E_{CBM}}{\partial V} \right)_T\right|_{at\ V=V_0}
\end{equation}
where $V$ is the volume, $E_{CBM}$ is the energy of the CBM and $V_0$ is the volume of the relaxed structure (i.e., zero pressure)\cite{carrier79,carrier80}. It should be noted that since the absolute value of $E_{CBM}$ is a function of the volume itself, we use the difference between the energy of the first conduction band and the first valence (core) band. Furthermore, the elastic and piezoelectric constants have been already calculated \emph{ab initio} for GaAs and InN, and are available in the literature. For GaAs, we use the values calculated by Beya-Wakata et al.\cite{carrier91}, and for InN we use the values calculated by Sarasamak et al.\cite{carrier89}, to obtain the piezoelectric coefficient and elastic constant used in the equations for piezoelectric scattering in Table \ref{tab:elastic_scattering_expressions}.

As a comparison, the electrical conductivity and Seebeck coefficient are also computed using the widely-used BTE-cRTA formulation. We choose the BoltzTraP package \cite{50}, which uses Fourier interpolation of the calculated bands, and differentiate the band energies to find the group velocities of the electrons. Other than the need to fit the relaxation time to experimental measurements of the conductivity, the BoltzTraP/BTE-cRTA implementation represents an otherwise parameter-free model that can be adapted to different semiconductor materials.

\section{Computational Methodology}
\label{abinitio}

For each semiconductor material, the geometry of the unit cell is optimized, and the density of states and band structure are calculated.  In the case of zinc blende GaAs and wurtzite InN, the unit cells are optimized using Kohn-Sham density functional theory (KS-DFT) \cite{hohenberg_physrev_1964,kohn_physrev_1965}, as implemented in VASP.  The generalized gradient approximation of Perdew, Burke, and Ernzerhof (GGA-PBE) \cite{gga1,gga2} is used to express the exchange-correlation potential, and  Projector Augmented Wave (PAW) potentials \cite{paw1,paw2} are used to represent the valence wavefunctions. Information regarding the structure of these two systems and their changes upon geometry optimization have been summarized in Table \ref{tab:geometry}. The initial structures are obtained \textcolor{black}{from the literature \cite{InN_structure,GaAs_structure}.}

\begin{table}
	\caption{\label{tab:geometry}Structure of GaAs and InN calculated with DFT, using the GGA-PBE exchange-correlation functional. Changes in the lattice constants compared to experimental values \cite{InN_structure,GaAs_structure} upon optimization are reported below.}
	\begin{ruledtabular}
	\begin{tabular}{cccccc}
		 Compound & Space Group & $\left| \mathbf{a} \right|$ (\AA) & \% change in $\left| \mathbf{a} \right|$ & $\left| \mathbf{c} \right|$ (\AA) & \% change in $\left| \mathbf{c} \right|$\\
		\hline
		GaAs & F-43m & 5.75 & 2.17\% & - & -   \\
		InN & P6$_3$mc & 3.533 & 0.56\% & 5.693 & 0.8\%\\
	\end{tabular}
\end{ruledtabular}
\end{table}

We then compute the electronic band structure of these materials. The energy cutoff for the plane wave basis set is set to $500 \ \textrm{eV}$. The band structure is computed in line mode along seven high-symmetry $k$-points in the IBZ, with 20 $k$-points between each pair of high-symmetry points.  The self-consistent density of states (DOS) calculation is performed using a $20 \times 20 \times 20$ $k$-point mesh, for both GaAs and InN. The non-self consistent energy calculations are performed in a special k-point mesh around the $\Gamma$ point, at which the conduction band minimum (CBM) occurs in both direct band gap GaAs and InN. This $k$-point mesh contains a total of 10,234 points in the Irreducible Brillouin Zone (IBZ), with mesh spacing of 0.001, 0.01, or 0.1 fractional units, to completely account for band anisotropy while remaining dense enough around the $\Gamma$ point to obtain accurate group velocity and effective mass values. To determine the effect of presumably more accurate band structure calculations on the band curvature, effective mass, and group velocity, we have also employed the $GW$ method. Only 941 $k$-points in the IBZ have been used for GW calculations, since this method is more computationally demanding. Using more $k$-points does not change the calculated effective mass. The $GW0$ band structure calculations are performed using the maximally-localized Wannier functions (MLWFs) interpolation, as implemented in VASP and the Wannier code \cite{93}. It is very important to also show the feasibility of the \emph{ab initio} model with band structures calculated using DFT+U, since the GW and hybrid functional (e.g., HSE\cite{hse,hse0,hse1}) methods are computationally demanding for complex materials with larger unit cells than GaAs. On the other hand, for InN, all methods and functionals attempted, including LDA, GGA, HSE and GW0, resulted in a band structure with zero band gap and a falsely predicted p-like conduction band. Only GGA+U \cite{hubbard}, with U values obtained from the literature \cite{carrier78,carrier77} (U$_d=6$ for In and U$_p=1.5$ for N), produced a correct band structure and with a more s-like conduction band particularly around $\Gamma$ point, which is consistent with the self-interaction corrected band structure reported by Furthm{\"u}ller et al. \cite{carrier14}. In the process of choosing the U value for GGA+U band structure calculations on GaAs, however, the values (U = 8 eV for both $d$ orbitals of Ga and As) recommended by Persson and Mirbt \cite{carrier76}, with an emphasis on correctly obtaining the band gap and effective mass values, result in GaAs falsely becoming an indirect semiconductor, with the conduction band minima located at the $L$ and $X$ $k$-points rather than the $\Gamma$ point \cite{carrier76}. Therefore, we have also employed effective U values of 7 eV (Ga) and 6 eV (As), for which a direct band structure is obtained. We have calculated mobilities obtained from both of these band structures and compared them with the ones obtained by the GW band structure. In order to calculate the group velocities, $v\left(k\right)$, and the overall average effective mass, we have fitted a sixth degree polynomial to the calculated conduction band (i.e., average energy as a function of distance from $\Gamma$ point or $\varepsilon\left(k\right)$) with $R^2 > 0.99$: 
\begin{equation}
v\left(k\right)=\frac{1}{\hbar}\frac{\partial \varepsilon}{\partial k}
\label{group_velocity}
\end{equation}
\begin{equation}
m^\ast=\left.\left(\frac{1}{\hbar^2}\frac{\partial^2 \varepsilon}{\partial k^2}\right)^{-1}\ \right|_{at\ k=0}
\label{effective_mass}
\end{equation}
It should be noted that we do not use the value of effective mass in the proposed carrier transport model. Rather, we calculate it solely to compare with experiment and evaluate the effect of the shape of the conduction band (i.e., group velocities) calculated by various methods, such as GGA, GGA+U, and GW. \textcolor{black}{Fitting polynomials to the numerically calculated conduction band and density of states results in smooth plots of mobility and Seebeck coefficient, as presented here, while preserving the values that are calculated \emph{ab initio} with $R^2 > 0.99$ in all segments fitted. We fit these polynomials at different segments of the band structure and carefully choose only the ones that result in the maximum $R^2$ and minimum discontinuity where the polynomials meet. This results in very smooth calculated group velocities, and, subsequently, other transport properties.}

\section{Results and Discussions}
\label{results}


\subsection{\emph{ab initio} Calculated Parameter Inputs to the Transport Model}

\begin{figure}
	\subfloat[GaAs]{
		\label{fig:BSDOS_GaAs}
		\includegraphics[width=0.475\columnwidth]{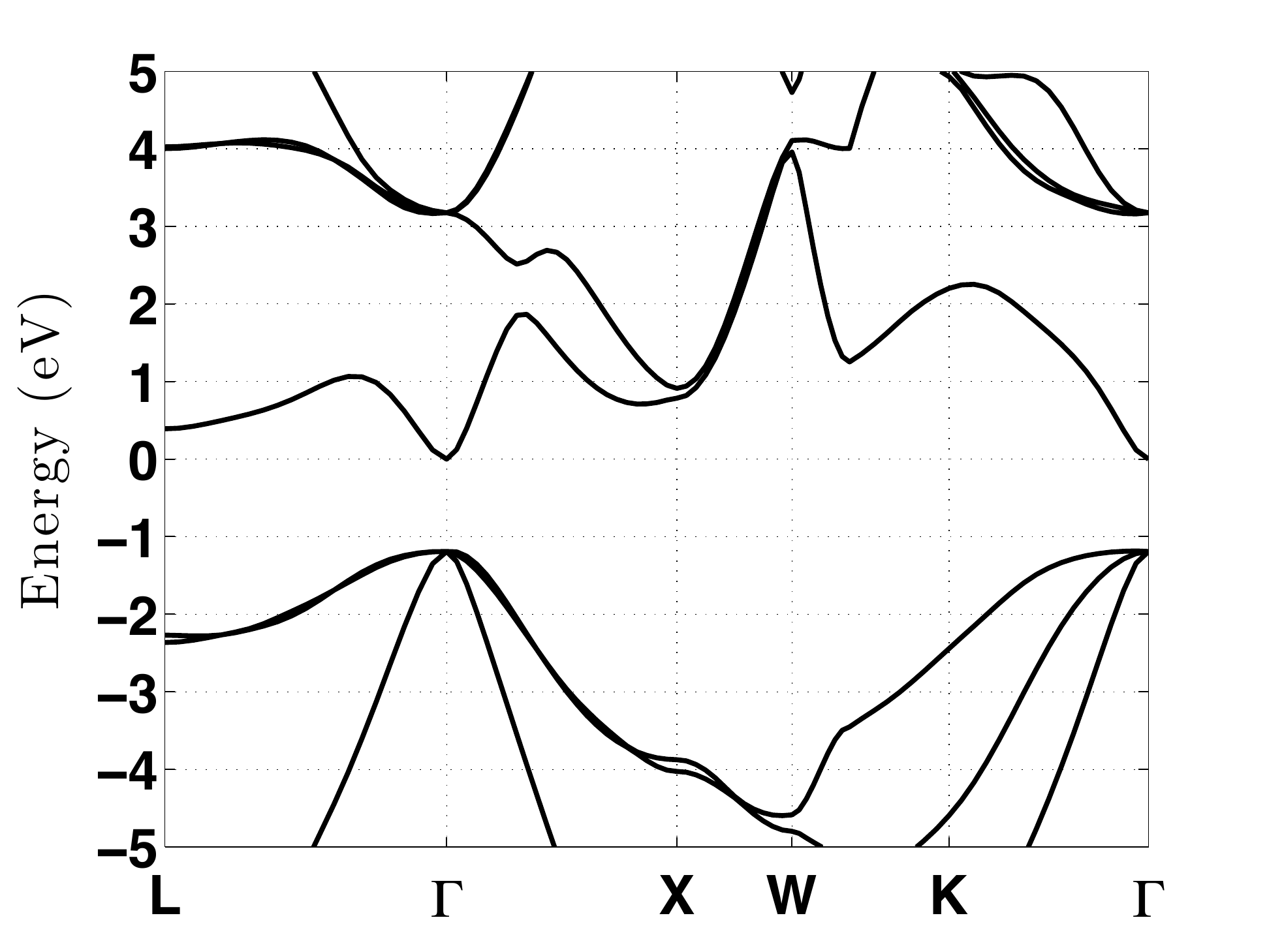}}
	\subfloat[InN]{
		\label{fig:BSDOS_InN}
		\includegraphics[width=0.475\columnwidth]{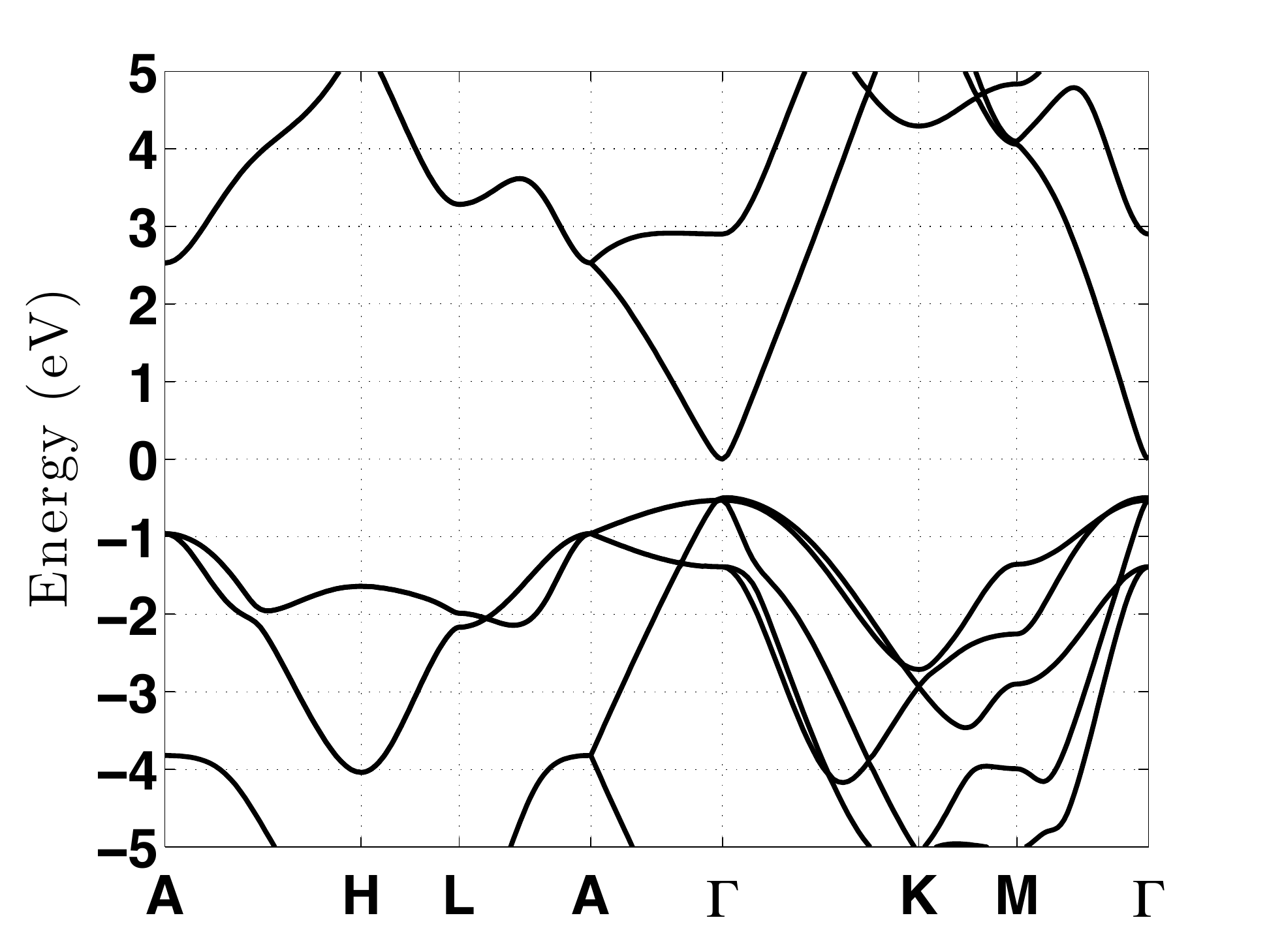}}
	\caption{Band structure of cubic GaAs and wurtzite InN, normalized so that the Fermi level is set to zero at the conduction band minimum.}
	\label{fig:BSDOS}
\end{figure}

The computed band structures of GaAs and InN are shown in Figure \ref{fig:BSDOS}. We have calculated a $GW0$ band structure, which starts from the wavefunctions previously computed using the GGA-PBE functional, as shown in Figure \ref{fig:BSDOS_GaAs}. 

\begin{table}
	\caption{\label{tab:inputs}Inputs to the transport model, as calculated \emph{ab initio} compared to experimentally measured values. The bolded numbers are used in our transport property calculations; note that not all appear in the final expressions listed in Table \ref{tab:elastic_scattering_expressions}.}
	\begin{ruledtabular}
		\begin{tabular}{lcccc}
			\multicolumn{1}{c}{}&  \multicolumn{2}{c}{GaAs} & \multicolumn{2}{c}{InN} \\ \cline{2-3} \cline{4-5}
			Parameter \footnotemark[1] & \emph{ab initio}  & Exp. & \emph{ab initio}  & Exp. \\ \hline
			$c_l$ (nm) &  \textbf{0.562} & 0.575  \ \cite{book8} & \textbf{0.565}  & 0.569 \ \cite{book8} \\
			$\omega_{po}$ (THz)  & \textbf{8.16}  & 8.73 \ \cite{book8} & \textbf{17.83} \ \cite{book8} & 17.65 \\
			$\epsilon_0$  & \textbf{12.18} & 12.91 \ \cite{book8} & \textbf{11.42} & 10.3 \ \cite{Ager_InN}  \\ 
			$\epsilon_\infty$ & \textbf{10.32} &  10.91 \ \cite{book8} & \textbf{6.24} &  6.7 \ \cite{Ager_InN} \\
			$E_D$ (eV) & \textbf{6.04} &  8.6 \ \cite{book8} & \textbf{4.46} &  3.6 \ \cite{book8} \\
			$m^\ast$ & 0.053-0.066 \footnotemark[2] &  0.0636-0.082 \ \cite{carrier11,carrier54} & 0.062, 0.071\ (GW) &  0.05-0.08 \ \cite{carrier43,carrier51,carrier52,carrier55,carrier57,carrier58}  \\
			$\varepsilon_g$ (eV)  & 0.96, 1.19 \ (GW) &  1.424 \ \cite{carrier26} & 0.50 & 0.675-0.7 \ \cite{carrier52,carrier43,carrier59}  \\	
		\end{tabular} 
	\end{ruledtabular}
	\footnotetext[1]{The parameters are: $c_l$ (lattice constant), $\omega_{po}$ (PO phonon frequency), $\epsilon_0$ (low-frequency dielectric constant), $\epsilon_\infty$ (high-frequency dielectric constant), $E_D$ (deformation potential), $m^\ast$ (effective mass), $\varepsilon_g$ (band gap).}
	\footnotetext[2]{The GaAs effective masses are calculated as 0.053 (GGA+U, this work), 0.066 (GGA+U, with published $U$ \cite{carrier76}), and 0.063 (GW0).}
\end{table}

\begin{figure}
			\subfloat[GaAs]{
		\label{fig:bands_GaAs}
		\includegraphics[width=0.475\columnwidth]{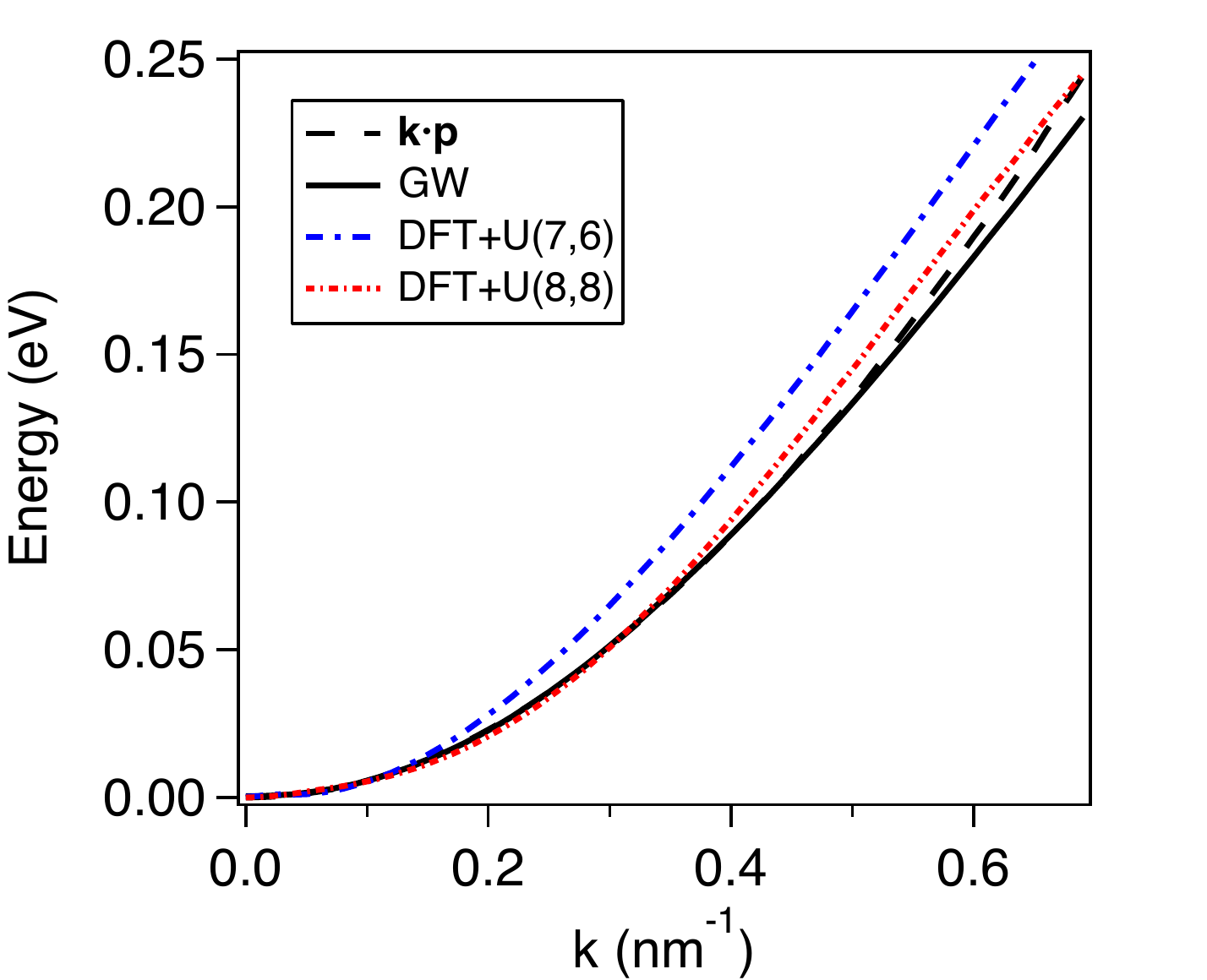}}
	\subfloat[InN]{
	\label{fig:bands_InN}
		\includegraphics[width=0.465\columnwidth]{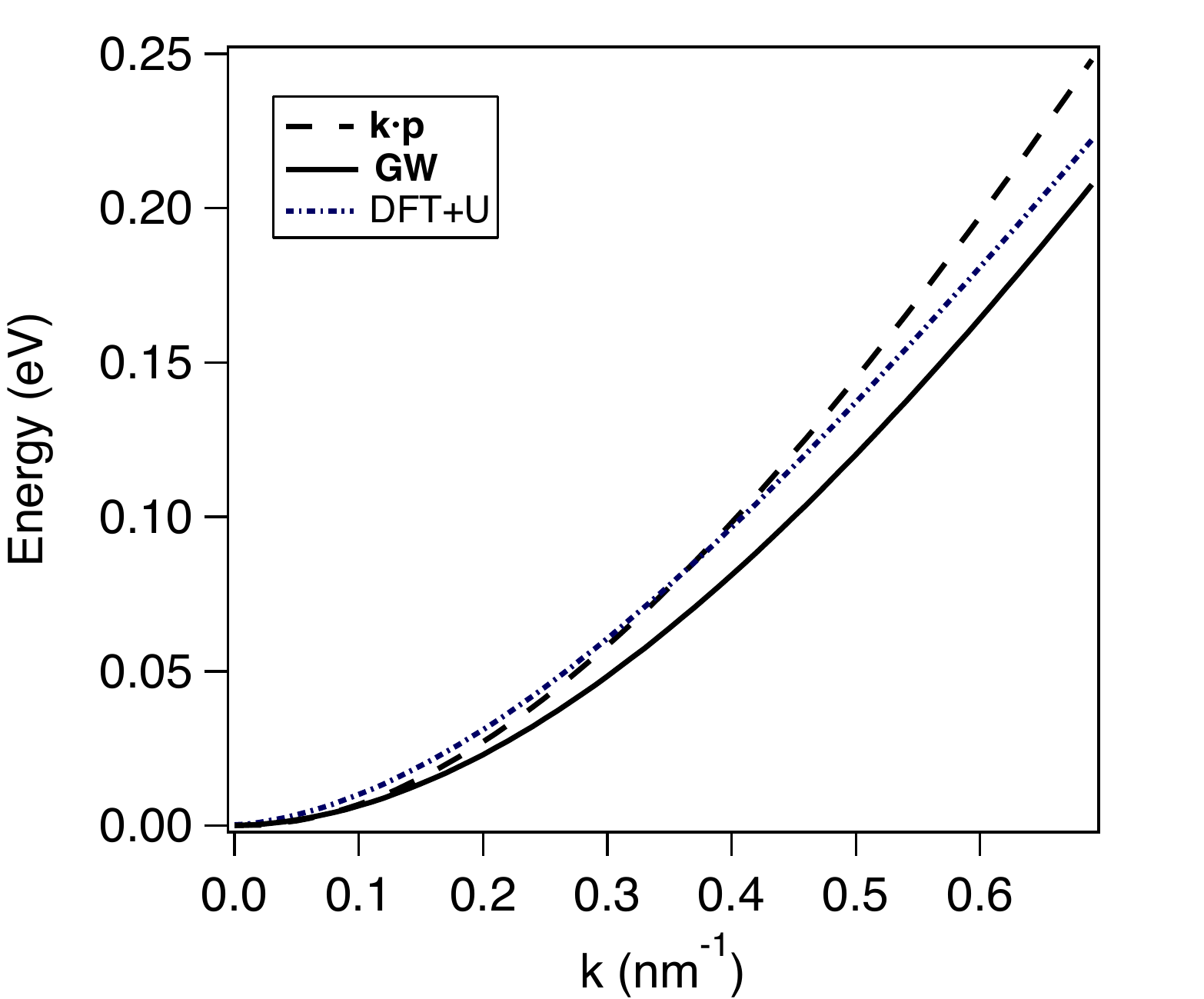}}
	\caption{The conduction bands expressed in terms of the average energy as a function of distance from the CBM (i.e., center of Brilloun zone, or $\Gamma$ point), as calculated from semi-empirical expressions (in $\mathbf{k}\cdot\mathbf{p}$ formulation) versus \emph{ab initio}. The difference at higher $k$ values has a significant impact on transport properties, especially at high temperatures. The values of U for the $d$ orbitals of Gallium and Arsenic, respectively, are in parentheses, while those for InN are taken from the published literature \cite{carrier78,carrier77}.}
\label{fig:bands1}
\end{figure}

The band structures used in previous semi-empirical models \cite{Ager_InN,book8} express the energy of the conduction band as a function of the distance from the $\Gamma$ point. Instead, we calculate the \emph{ab initio} band structure in a three-dimensional grid around the CBM, and then average the energy values of the $k$-points that share the same distance from the $\Gamma$ point (Figure \ref{fig:bands1}).  For both GaAs and InN, the \emph{ab initio} and $\mathbf{k}\cdot\mathbf{p}$ band structures agree well at small $k$-points; however, they diverge at larger $k$-points.  This directly impacts the group velocity of the electrons and, ultimately, the transport properties -- particularly at higher temperatures where higher energy electrons have nonzero occupation.

We have \textcolor{black}{also} calculated a GGA+U \cite{hubbard} band structure, with U values taken from the published literature \cite{carrier78,carrier77}, as shown in Figure \ref{fig:BSDOS}. 
For InN, GGA+U correctly yields an s-like conduction band and a band gap of 0.5 eV, which is comparable to the self-interaction corrected band gap of 0.58 eV reported by Furthm{\"u}ller et al. \cite{carrier14} and the experimental values of 0.675-0.7 eV \cite{carrier52,carrier43,carrier59} (Table \ref{tab:inputs}). 
\textcolor{black}{We include DFT+U calculations only to show the feasibility of these less-expensive methods, in the case of more complex semiconductor materials for which a $GW$ calculations is too expensive. Also, DFT usually suffers from vastly underestimating the effective mass\cite{carrier24,carrier25}, and the introduction of the fitting parameter U may reduce the predictability of the \emph{ab initio} model as a whole. Therefore, we stress that all reported transport properties are calculated here using the parameter-free $GW$ band structures, unless otherwise stated.}

Although we do not directly use the value of the electron effective mass in the transport property expressions, we see that the calculated effective mass of 0.062 for InN is consistent with the previously calculated effective mass (0.066) using an empirical pseudopotential \cite{carrier10}, and well within the range (0.05-0.08) measured experimentally \cite{carrier52,carrier43,carrier51,carrier55,carrier57,carrier58}. 

We also show the calculated phonon band structure and density of states of these two compounds in Figure \ref{fig:phonon}. For GaAs, the calculated PO-phonon frequency of 8.16 THz is shown in Figure \ref{fig:phonon_GaAs}. For InN, the calculated optical phonon frequency of 17.83 THz is close to the 17.65 THz value reported by Bungaro et al. \cite{carrier15,carrier43}. \textcolor{black}{We have listed all the parameters that are used in our transport model in Table \ref{tab:inputs}. We have calculated all of these parameters, as bolded in Table \ref{tab:inputs}, \emph{ab initio} to demonstrate the feasibility of a fully predictive model for transport properties. The only exceptions are the elastic and piezoelectric constants, which are necessary to calculate the piezoelectric coefficient, $P$, in Table \ref{tab:elastic_scattering_expressions}. As described earlier, we have instead used the previously calculated values from published DFT studies for these constants\cite{carrier91,carrier89}}.

\subsection{Model Validation on GaAs}

In order to evaluate the accuracy of our model, we first calculate the mobility of three experimentally synthesized and characterized GaAs samples, as described by Stillman et al. \cite{carrier20}. We also perform this analysis over a wide temperature range for high purity GaAs samples with very low electron concentrations, as labeled as "pure" in Table \ref{tab:GaAs_samples}.

\begin{table}
	\caption{\label{tab:GaAs_samples}Carrier concentrations of various experimentally fabricated and characterized GaAs samples. For the "pure" sample, data is available roughly between 5-1000 K.  For the real samples, mobility data is also tabulated at different temperatures.}
	\begin{ruledtabular}
	\begin{tabular}{ccccc}
		 Sample & Concentration, $n  \ \left(\text{cm}^{-3} \right)$ & Donor, $N_D$ & Acceptor, $N_A$& Reference\\ \hline
		 pure			&	$3 \times 10^{13}$ & $5.2 \times 10^{13}$ & $2.2 \times 10^{13}$& \cite{book8} \\ 
		 a 		&	$ 2.7 \times 10^{13}$ & $4.8 \times 10^{13}$& $2.1 \times 10^{13}$&	 \cite{carrier20} \\
		 c			&  $7.7 \times 10^{14}$ & $1.1 \times 10^{15}$& $3.3 \times 10^{14}$&	\cite{carrier20} \\
		 e			&	$3.1 \times 10^{15} $ & $4.7 \times 10^{15}$ & $1.6 \times 10^{15}$&	\cite{carrier20} \\ 
		
	\end{tabular}
\end{ruledtabular}	
\end{table}

\begin{figure}
			\subfloat[Comparison of model trends with varying temperature and electron concentration to experimental data.  The values of U used for the $d$-orbitals of Gallium and Arsenic, respectively, are in parentheses.]{
		\label{fig:GaAs_samples}
		\includegraphics[width=0.475\columnwidth]{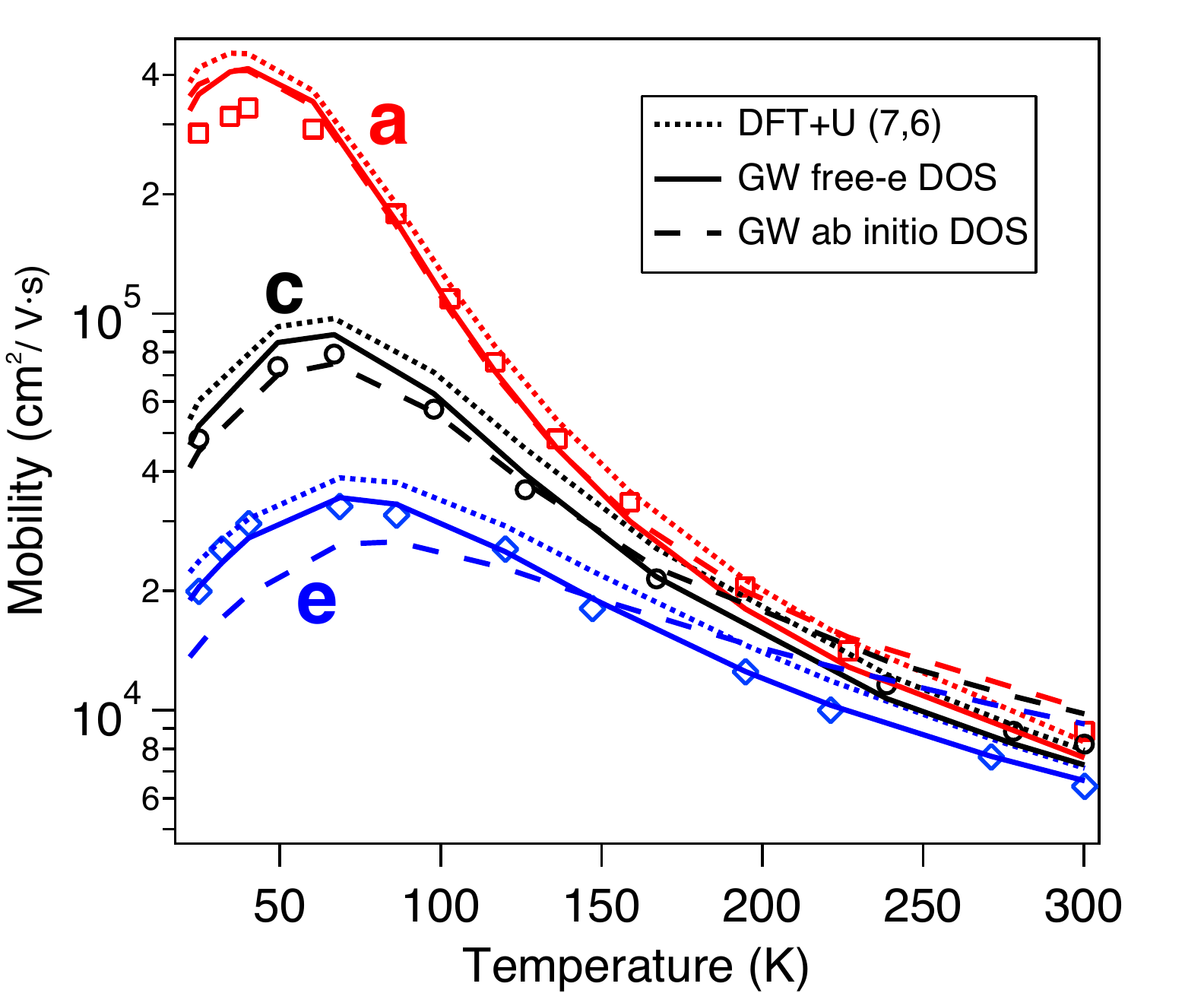}}
	\subfloat[Limitation of "pure" GaAs mobility from each scattering mechanism.]{
	\label{fig:GaAs_lowT}
		\includegraphics[width=0.475\columnwidth]{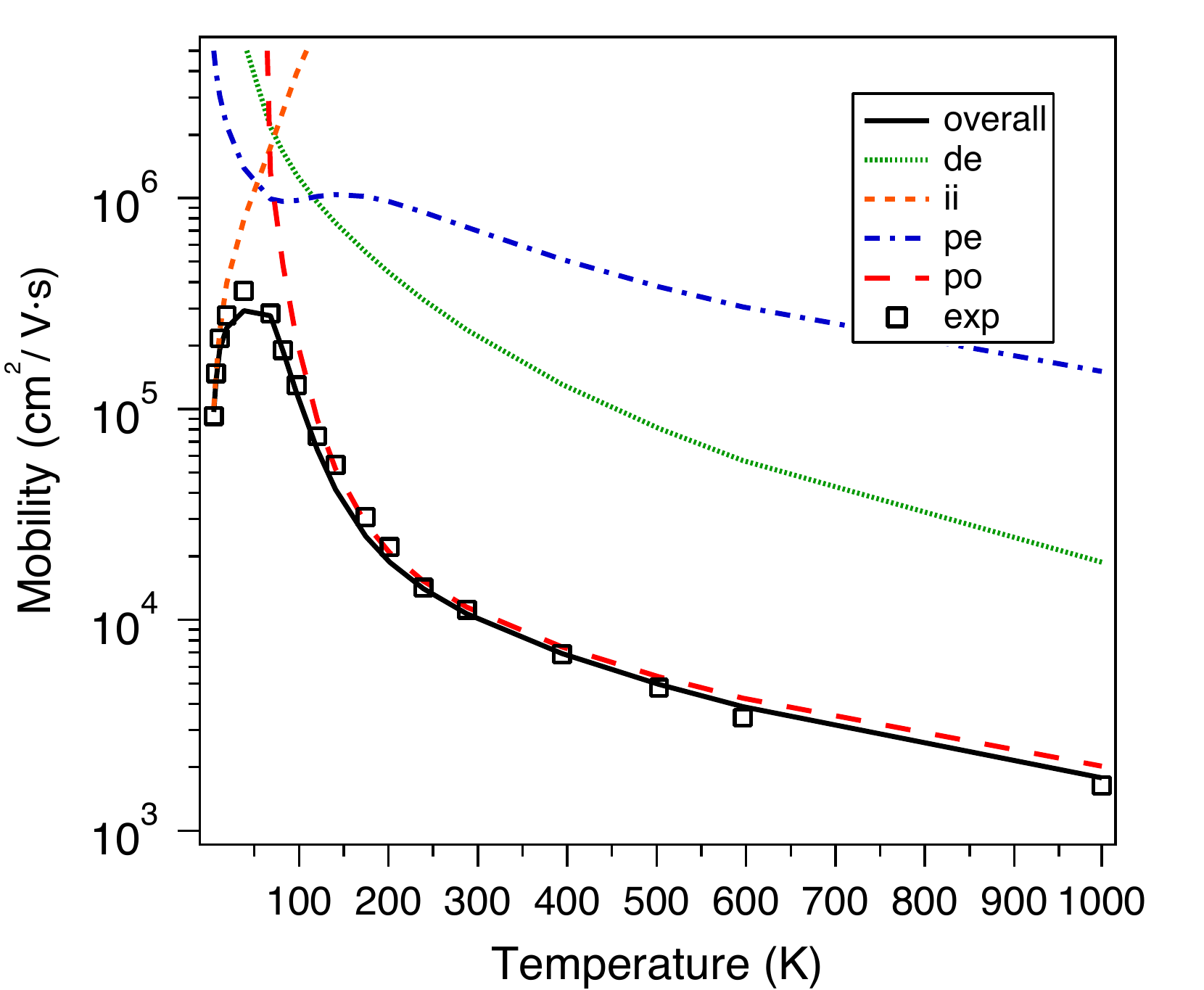}}
	\caption{The calculated and experimental \cite{book8,carrier2} mobility data for GaAs at various electron concentrations and temperatures. More details on the experimental data, including donor and acceptor concentrations, are available in Table \ref{tab:GaAs_samples}.}
\end{figure}

\textcolor{black}{As shown in Figure \ref{fig:GaAs_samples}, the most accurate $GW$ band structure results in the best agreement with experimental data.  The DFT+U band structure, however, does provide us with limits of the mobility over different temperatures. When calculating the mobility and Seebeck coefficient, we calculate the Fermi level by first calculating the electron concentration through Equation \ref{eq:e-concentration}, and then matching it to a given concentration. The calculated properties are very sensitive to the calculated Fermi level. Therefore, for comparison, we have included the results using both the \emph{ab initio} DOS used in Equation \ref{eq:e-concentration}, and the free electron DOS. As shown in Figure \ref{fig:GaAs_samples}, the \emph{ab initio} model for DOS performs better for lower electron concentrations and lower temperatures, while the free electron DOS is more suitable for higher temperatures, and, particularly, at higher electron concentrations. We acknowledge that because of the log scale in Figure \ref{fig:GaAs_samples}, seeing the quantitative agreement is difficult. Therefore, we report the calculated relative error compared to the experiment for the best cases for each sample -- from the \emph{ab initio} DOS for sample \textbf{a} and from the free electron DOS for samples \textbf{c} and \textbf{e}. The minimum, maximum and the relative error in calculating the mobility of sample \textbf{a} are 2.25\% (at 195 K), 29.42\% (at 29 K), and 13.33\%, respectively. These numbers are 1.02\% (at 167K), 15.01\% (at 49K), and 7.97\% for sample \textbf{c} and  0.22\% (at 195K), 7.90\% (at 40K), and 4.04\% for sample \textbf{e}.} Overall, the agreement is poorer at higher electron concentrations and lower temperatures; this is attributed to the inaccuracy of the Brooks-Herring ionized impurity scattering model at high electron concentrations, as briefly described in Section \ref{theory}. Furthermore, the model has also been validated with the data on crystalline samples with very high purity. \textcolor{black}{The calculated electron mobilities, assuming the limit that only one scattering mechanism exists at a time, along with the overall mobility, are shown in Figure \ref{fig:GaAs_lowT}. The reasonable agreement between the calculated and experimental mobilities provides independent validation of the transport model. The minimum, maximum and average relative error of calculated mobility are 0.46\% (at 394K), 23.55\% (at 175K) and 9.53\% respectively for temperatures above 20 K. }
The mobility is mainly limited by ionized impurity scattering at low temperatures, piezoelectric scattering at intermediate temperatures, and polar optical phonon scattering at higher temperatures ($> 60 \ \text{K}$); all of these are consistent with the previous results shown by semi-empirical models \cite{book8,carrier26,carrier32} \textcolor{black}{yet no experimental parameter has been used here in predicting the correct changes with the temperature and the carrier concentration.}


\begin{figure}
		\includegraphics[width=0.475\columnwidth]{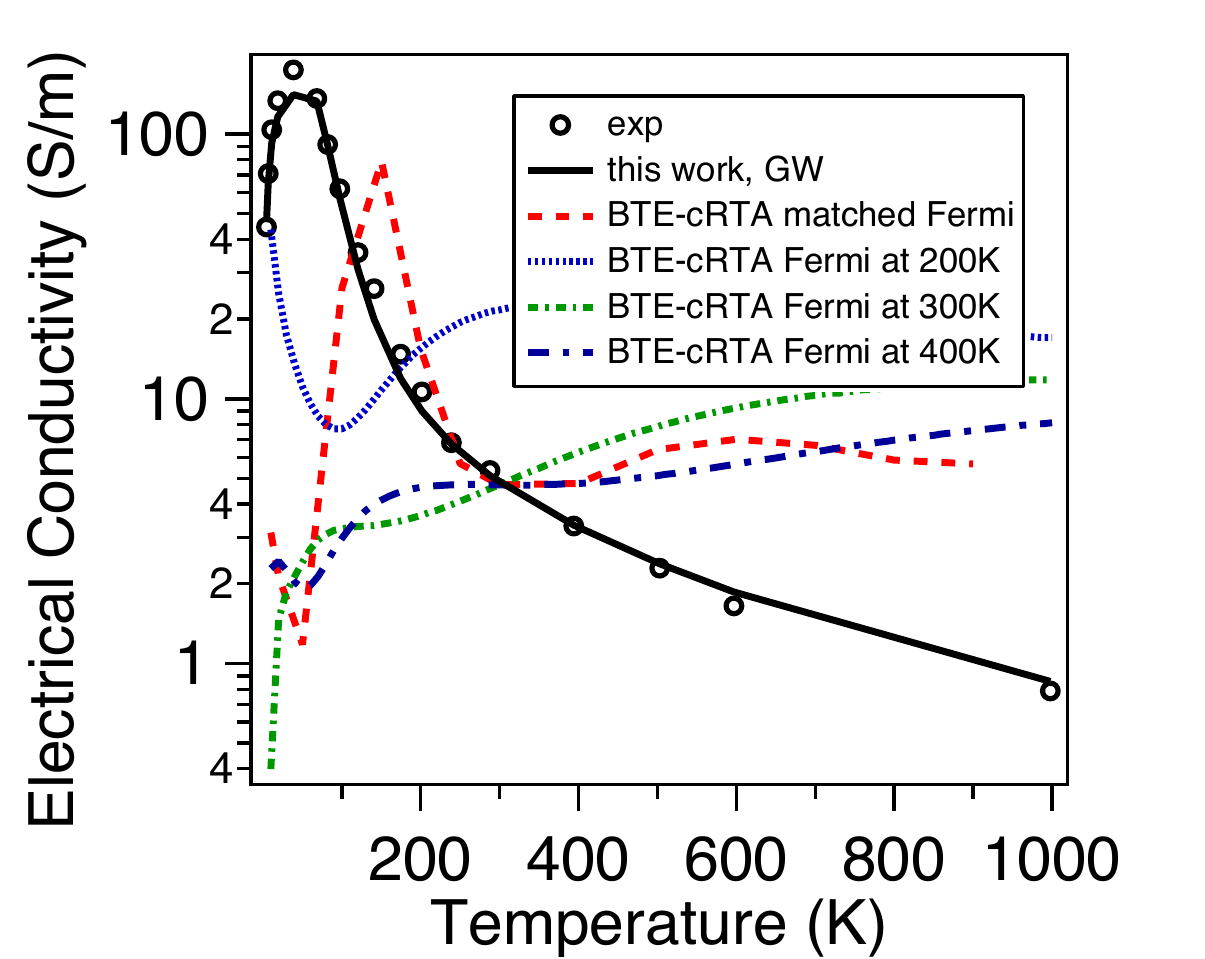}
	\caption{Electrical conductivity of GaAs calculated using the \emph{ab initio} transport model and the BTE-cRTA framework, and compared to experimental \cite{book8} data. The Fermi level is calculated by matching the calculated carrier concentration to $ n = 3 \times 10^{13}$. This has been done either at the mentioned temperature and kept constant over the whole temperature range, or in the case of "matched Fermi", at each temperature, the Fermi level is adjusted to the given $n$. The relaxation time, $\tau$, is determined by fitting the calculated conductivity to the corresponding experimental value at 300 K. The calculated value of $4.5 \times 10^{-23} s $ for $\tau$ is unreasonably low but it has been included for the sake of comparison of all models. }
			\label{fig:GaAs_BoltzTraP}
\end{figure}

Once we have the calculated mobility, at a given electron concentration, we can calculate the electrical conductivity of GaAs by Equation \ref{eq:conductivity_simple}. For now, we assume that the carrier concentration remains constant with temperature over the range of interest. We then compare to the experimental conductivity and those values calculated using the BTE-cRTA framework, under the scenarios listed in Figure \ref{fig:GaAs_BoltzTraP}. As shown, not only does BTE-cRTA fail to correctly predict the trend for conductivity with temperature, but also quantitatively differs from the experimental values.

\begin{figure}
	\includegraphics[width=0.475\columnwidth]{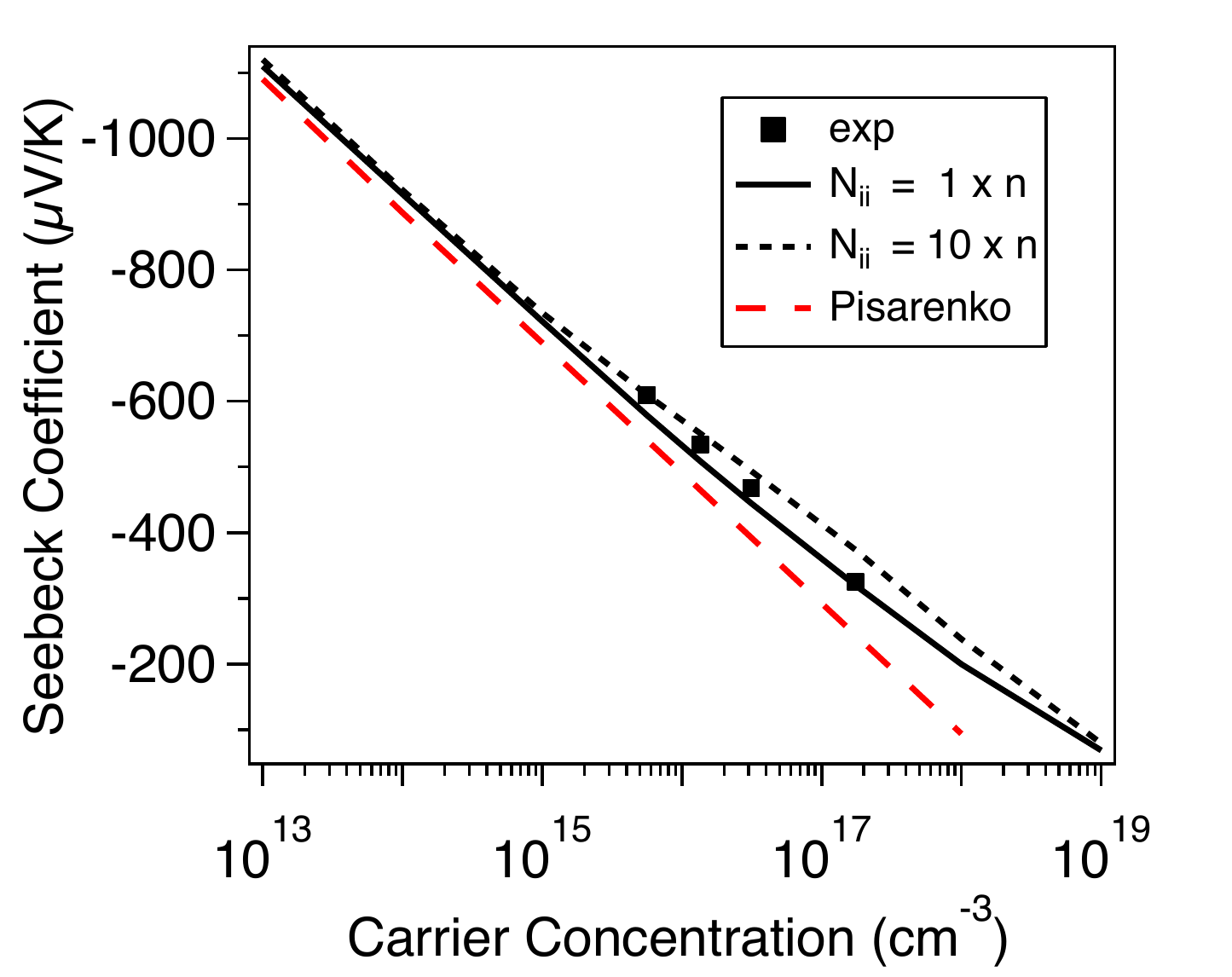}
	\caption{Calculated and experimental \cite{carrier32} GaAs Seebeck coefficient, at different ratios of the ionized impurity concentration, $N_{ii}$, to the electron concentration, $n$. \textcolor{black}{For the Pisarenko plot, using Equation \ref{eq:pisarenko}, we have used values of $r=-\frac{1}{2}$ for acoustic phonons, and $m^\ast = 0.063$.}  }
	\label{fig:GaAs-s}
\end{figure}

Finally, we calculate the Seebeck coefficients of the GaAs samples (assumed to be at 300 K), and compare them to the values reported previously by Rode and Knight \cite{carrier32} (Figure \ref{fig:GaAs-s}). Since the data are for various samples with different electron concentration and compensation ratios, we choose various values of $N_{ii}/n = (N_D+N_A)/n$. As shown, a range of Seebeck coefficients are calculated at each electron concentration, which includes the experimentally measured points. It should be noted that not knowing beforehand the compensation and concentration of donors and acceptors, as well as their charge states, limits the overall predictability of our model. However, even given these limitations, the close fit between \emph{ab initio} and experimental properties provides independent validation of the viability of our model. \textcolor{black}{For further evaluation, we have calculated the Seebeck coefficient, assuming Pisarenko behavior and compared it to our model in Figure \ref{fig:GaAs-s}. We use Equation \ref{eq:pisarenko} with two fitting parameters: effective mass, $m^\ast$ and $r$.  It should be noted that in the case where the best agreement with experiment, through Pisarenko behavior, is only achievable by choosing either $m^\ast = 0.11$ or $r = 0.35$, both of these values are far from experimental measurements and thus lack physical meaning.}

\begin{equation}
S \simeq \frac{k_B}{e}\left[\frac{5}{2} + r + ln\frac{2\left(2\pi m^\ast k_BT\right)^{3/2}}{h^3n}\right]
\label{eq:pisarenko}
\end{equation}

\subsection{Model Validation on InN}

\begin{figure}
	\subfloat[Mobility]{
		\label{fig:samples_InN-mu}
		\includegraphics[width=0.475\columnwidth]{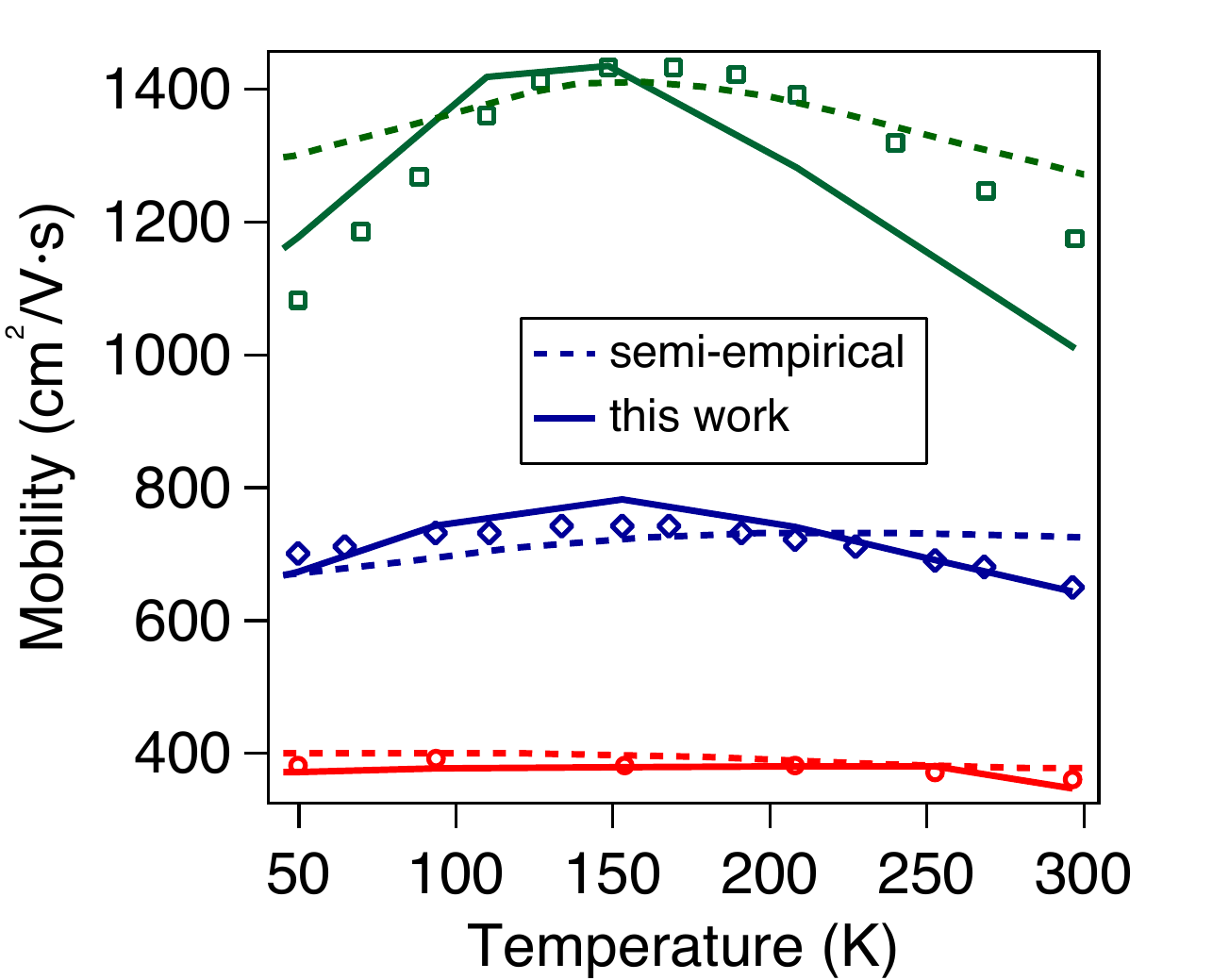}}
	\subfloat[Seebeck coefficient]{
		\label{fig:samples_InN-s}
		\includegraphics[width=0.475\columnwidth]{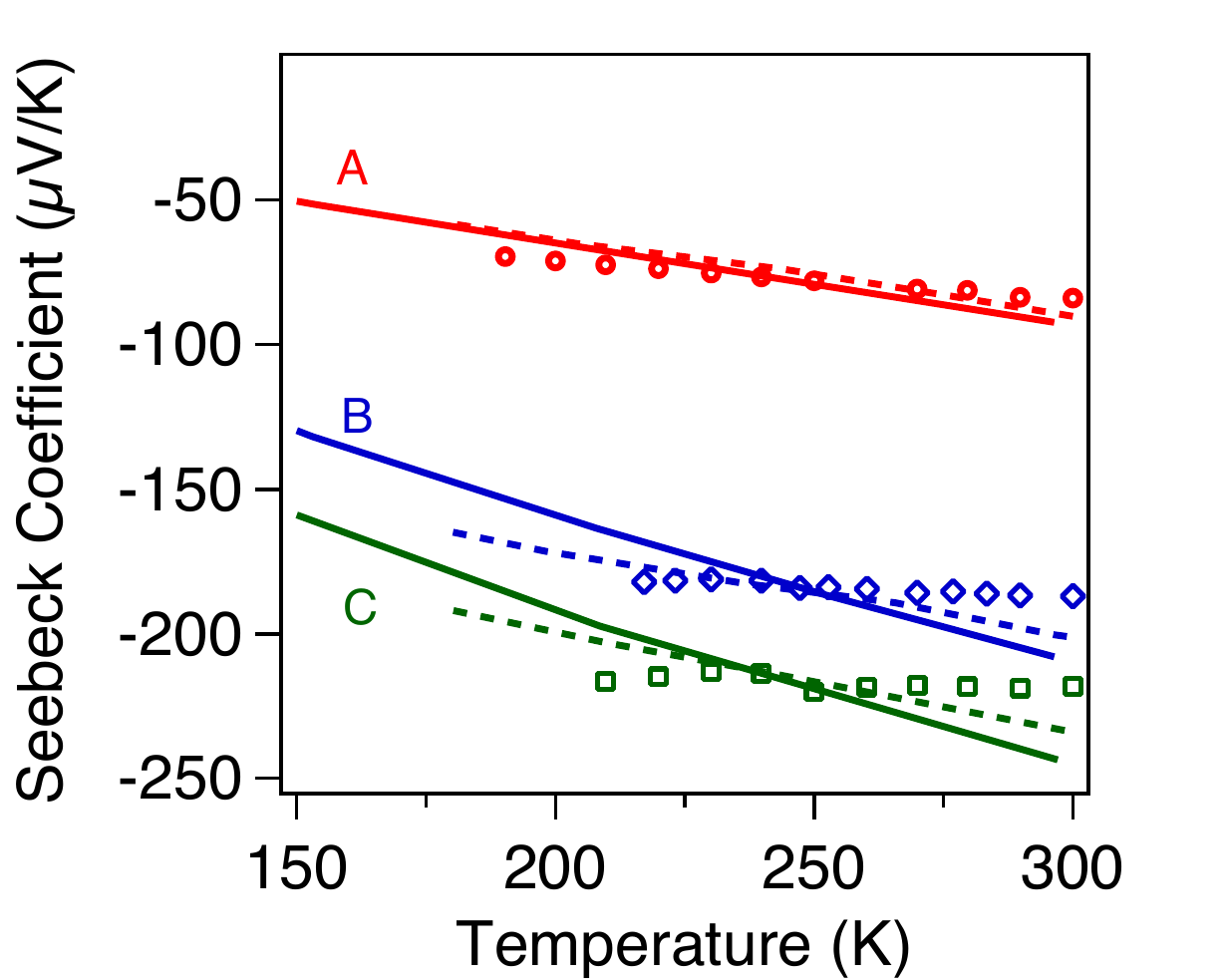}}
	\caption{Calculated and experimental transport properties of the three InN samples listed in Table \ref{tab:InN}. The dashed line is calculated by the semi-empirical model used by Miller et al. \cite{Ager_InN}, while the solid line is calculated by the proposed \emph{ab initio} transport model.}
	\label{fig:samples_InN}
\end{figure}

In order to further evaluate the accuracy of our model and its applicability to more complicated semiconductors, we also calculate the mobility and Seebeck coefficient (Figure \ref{fig:samples_InN}) of three experimentally synthesized and characterized InN samples by Miller et al. \cite{Ager_InN}.  These calculations are more challenging due to the reported presence of linear charged dislocations in the crystal structure \cite{Ager_InN,dislocations1,dislocations2,dislocations3}, due to the processing conditions employed. For each sample at a given carrier concentration, as shown in Table \ref{tab:InN}, we change the concentration of dislocations, $N_{dis}$, until the calculated mobility values match the experimental measurements. The fitted $N_{dis}$ (Table \ref{tab:InN}) is within the range of measured concentrations from transmission electron microscopy analysis (TEM) \cite{Ager_InN}, which confirms that the limiting mechanism is indeed scattering from dislocation lines.

\begin{table}
	\caption{\label{tab:InN}Measured \cite{Ager_InN} and calculated InN dislocation density, corresponding to the mobility and Seebeck coefficient reported in Figures \ref{fig:samples_InN-mu} and \ref{fig:samples_InN-s}.}
\begin{ruledtabular}
	\begin{tabular}{cccc}
		& \multicolumn{3}{c}{$N_{dis} \ \left(\text{cm}^{-2} \right)$} \\
		\cline{2-4}
		 Sample & Experimental & Semi-empirical\cite{Ager_InN}& This work \\ \hline
		 A 		&	$\approx 1 \times 10^{11}$ & $1.5 \times 10^{11}$& $8.20 \times 10^{10}$\\
		 B			&  $2-5 \times 10^{10}$ & $1.5 \times 10^{10}$& $1.18 \times 10^{10}$\\
		 C			&	$\approx 1 \times 10^9-5 \times 10^{10}$ & $4.1 \times 10^{9}$&	$3.47 \times 10^{9}$\\ 
	\end{tabular}
	\end{ruledtabular}
\end{table}

\begin{figure}
		\includegraphics[width=0.475\columnwidth]{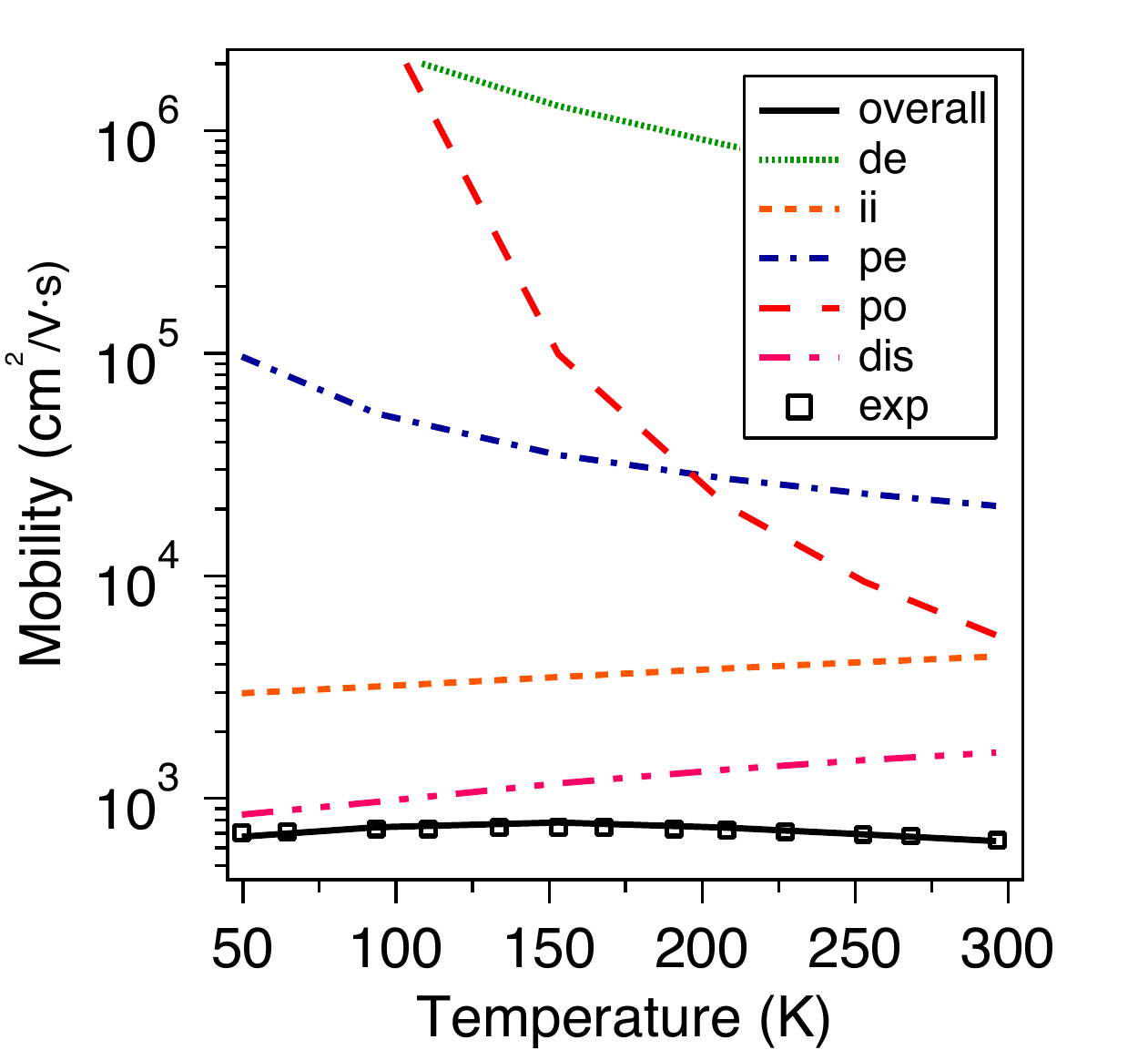}
	\caption{Calculated and experimental \cite{Ager_InN} values for InN mobility at $n = 9\times 10^{17} \ \text{cm}^{-3}$ (sample B in Table \ref{tab:InN}). Each line represents the mobility if limited only by the corresponding mechanism.}
		\label{fig:InN-decomposed}
\end{figure}

As shown in Figure \ref{fig:samples_InN-mu} and \ref{fig:samples_InN-s}, while there is an excellent agreement between the calculated and experimental mobility, the calculated Seebeck coefficients for samples B and C exhibit more pronounced changes with temperature than the experimental Seebeck coefficients. 
The mobility of the samples is found to be limited by charged dislocations, particularly at low temperatures. The next limiting mechanism is polar optical phonon scattering, which is more important at higher temperatures while ionized impurity scattering is more important at lower temperatures. This can be seen in Figure \ref{fig:InN-decomposed}, which shows the mobility of sample B if it were limited by each type of scattering mechanism, as well as the overall mobility. These findings are in agreement with the semi-empirical transport model \cite{Ager_InN}, except that all parameters are obtained from \emph{ab initio} calculations that require knowledge only of the crystal structure of the material. \textcolor{black}{Comparing the transport properties calculated from using model with those calculated using semi-empirical models (including experimentally measured band gap and effective mass (See Table \ref{tab:inputs} under "Exp.") in Figure \ref{fig:samples_InN} shows that although quantitative agreement with experiment is slightly better with the semi-empirical model, Seebeck coefficient calculations on samples B and C, and the mobility of the sample at high temperature, show much better accuracy with the \emph{ab initio} model presented here.}

Finally, we should once again acknowledge the assumptions and limitations of the current model when applied to the other types of semiconductors. Most importantly, the formulation presented in this work is for low-field transport (particularly drift mobility and Seebeck coefficient), in which the changes to the electron distribution are merely a linear perturbation to the equilibrium Fermi-Dirac distribution; thus, the applicability of the current model for high-field transport or heavily doped and polar semiconductors where the linear BTE formulation fails \cite{kohn_physrevb_1957}, is very limited. Furthermore, we have averaged the energy around CBM and expressed the energy values in the band structure as a function of the absolute value of $\textbf{k}$, or simply, the distance from $\Gamma$ point in the reciprocal space. Therefore, the reported mobility values are averaged and the effect of band structure anisotropy is not fully captured. It is possible, however, to include the band structure of the material only in the specific orientation of interest to account for anisotropy. \textcolor{black}{Currently, the model is limited to a single conduction band. Although the single band \emph{ab initio} model can be used for prediction of many direct band gap semiconductors, it will only result in an overestimation of transport properties of semiconductors with more complex band structure. This is due to the fact that currently, interband scatterings between several bands that are participating in transport are neglected. In future, we will solve coupled-BTE and take into account two and more participating bands which enables calculation of both electron and hole mobilities in more materials}. Finally, although the usage of the Hubbard $U$ parameter in the band structure calculation might limit the predictability of the model in calculating overall transport properties, this can be properly addressed by using more accurate methods of band structure calculations \textcolor{black}{as reported here. We include DFT+U calculations here only to show the feasibility of working with the model when $GW$ or other less commonly used methods are not technically or otherwise feasible.}

\section{Conclusions}
\label{conclusions}

We have presented an \emph{ab initio} transport model for calculating the electrical mobility and Seebeck coefficient of n-type semiconductors. By using the inputs from density functional theory calculations, and considering all relevant physical phenomena (i.e., elastic and inelastic scattering mechanisms), we have successfully calculated highly-accurate transport properties of GaAs and InN over various ranges of temperature and carrier concentration. Our model provides both qualitative and quantitative improvements in accuracy compared to the widely-used semi-empirical and constant relaxation time approximation model solutions to the Boltzmann transport equation.  Future work will focus on extending this model to p-type semiconductors.

\textbf{Acknowledgement}
This research is based upon work supported by the Solar Energy Research Institute for India and the U.S. (SERIIUS) funded jointly by the U.S. Department of Energy subcontract DE AC36-08G028308 (Office of Science, Office of Basic Energy Sciences, and Energy Efficiency and Renewable Energy, Solar Energy Technology Program, with support from the Office of International Affairs) and the Government of India subcontract IUSSTF/JCERDC-SERIIUS/2012 dated 22nd Nov. 2012. This work used the Extreme Science and Engineering Discovery Environment (XSEDE), which is supported by National Science Foundation grant number OCI-1053575. We would like to thank Daniel Rode for helpful discussions during the preparation of this manuscript.

\clearpage

\begin{appendix}

\section{Boltzmann Transport Equation}
The Boltzmann transport equation (BTE) describes the non-equilibrium behavior of charge carriers (e.g., electrons or holes) by statistically averaging over all possible quantum states.  For the electron distribution, $f$, this is represented by the BTE:
\begin{equation}
\label{eq:btegeneral}
\frac{df \left( \mathbf{k},T,t \right)}{dt} = \left( \frac{\partial f \left( \mathbf{k},T,t \right)}{\partial t} \right)_s - \frac{d \mathbf{k}}{dt} \cdot \nabla_k f \left(\mathbf{k},T,t \right)- \mathbf{v} \left(\mathbf{k} \right) \cdot\nabla_r f \left( \mathbf{k},T,t \right)
\end{equation}
where $f$ is a function of state $\mathbf{k}$, temperature $T$, and time $t$, and $\mathbf{v} \left( \mathbf{k} \right)$ are the electron group velocities.  The three terms on the right-hand side of Equation \ref{eq:btegeneral} refer, respectively, to the temporal rate of change of $f$ due to all scattering processes, rate of change of $f$ due to external forces, and diffusion from the carrier density gradient.  

If the external forces consist only of a low electric field, $\mathbf{E}$, and no magnetic field, $\mathbf{B}$, such that $\frac{d \mathbf{k}}{dt} = \frac{e \mathbf{E}}{\hbar}$, 
then the low-field BTE becomes:
\begin{equation}
\label{eq:btesmallE}
\frac{df \left( \mathbf{k},T,t \right)}{dt} + \mathbf{v} \left( \mathbf{k} \right) \cdot \nabla_r f \left( \mathbf{k}, T \right) + \frac{ e \mathbf{E}}{\hbar} \cdot \nabla_k f \left( \mathbf{k}, T \right) = \left( \frac{\partial f \left( \mathbf{k},T,t \right)}{\partial t} \right)_s
\end{equation}

\subsection{Constant Relaxation Time Approximation}

Furthermore, $f$ can be described as a first-order (linear) perturbation, $g \left( \mathbf{k} \right)$, from the (equilibrium) Fermi-Dirac distribution, $f_0$, due to scattering:
\begin{equation}
\label{eq:perturbation} \left( \frac{\partial f \left( \textbf{k}, T, t \right)}{\partial t} \right)_s = - \frac{ f \left( \mathbf{k} \right) - f_0 \left( \mathbf{k} \right)}{\tau} = - \frac{g \left( \textbf{k} \right)}{\tau} \\
\end{equation}
 \begin{equation}
\label{eq:Fermi-Dirac} 
f_0 \left[ \varepsilon \left( \mathbf{k} \right) \right] = \frac{1}{e^{\left[ \varepsilon \left( \mathbf{k} \right) - \varepsilon_F \right] / k_B T} + 1}
\end{equation}
where the dependence of $\varepsilon$ on $\mathbf{k}$ is given by the electronic band structure, 
and the various scattering terms and time dependence are lumped into the electronic relaxation time, $\tau$.

If $\tau$ is a constant, then this major simplification results in the BTE-cRTA. This assumption simplifies the theory to an extent that closed form expressions for conductivity and Seebeck coefficient can be obtained \cite{50}. In this approach, the details of all elastic and inelastic scattering mechanisms are lumped into the relaxation time constant, $\tau$.  While popular, this approach suffers from the following disadvantages: 1. $\tau$ is obtained by fitting to the experimental data for the conductivity of the material, which limits the predictability of the model, and 2. Due to oversimplification of the transport mechanism, it may result in incorrect values and even incorrect trends with temperature or carrier concentration, as illustrated in Figure \ref{fig:GaAs_BoltzTraP}. Therefore, by explicitly including all possible electronic scattering mechanisms, one can determine which mechanisms are physically relevant for a given semiconductor.

\subsection{Explicit Solution of Linear BTE}
\label{appendix:btesolution}

To go beyond the relaxation time approximation, both elastic scattering mechanisms, for which the kinetic energy of the electrons remains constant, and inelastic scattering mechanisms, for which there is a change in the electron distribution, should be taken into account. If the system is governed only by elastic scattering mechanisms, the relaxation time, $\tau$, is equal to the inverse of the overall elastic scattering rates, which is the sum of all individual rates. Evidently, $\tau$ is not constant but does depend on energy; however, it does not necessarily follow a power law dependence (e.g., in InN \cite{Ager_InN}). However inelastic scattering mechanisms also limit the mobility, and therefore, the conductivity, of the semiconductor; as an example, polar optical (PO) phonon scattering is the main electron-phonon interaction that limits mobility at high temperatures in GaAs.  Thus, we need to first calculate the perturbation, $g$, to the electron distribution due to elastic and inelastic scattering mechanisms, and then integrate $g$ over all states to obtain the mobility.  Details on this approach are given below.

The most relevant elastic scattering mechanism for compound semiconductors is expected to be ionized impurity scattering at low temperatures.  Ionized impurity scattering occurs when a charged center is introduced inside the bulk material.  As a result of Coulombic interactions between the electron and ion, electrons scatter to different states (i.e., become distracted).  The ionized impurity scattering rate, $\nu_{ii}$ (i.e., a component of the overall $\nu$), may be expressed using Brooks-Herring theory \cite{carrier33}:
\begin{equation}
\label{eq:iiscattering}
\nu_{ii}= \frac{e^4N}{8\pi\epsilon_0^2\hbar^2k^2v}\left[D \ln\left(1+\frac{4k^2}{\beta^2}\right)-B\right]
\end{equation}
where the charge screening potential, $\phi$, is obtained by solving Poisson's equation:
\begin{equation}
\label{eq:iiscatteringphi}\phi = \frac{q}{4 \pi \epsilon_0 r} \exp \left( - \beta r \right) 
\end{equation}
and inverse screening length, $\beta$, is given by:
\begin{equation}
\label{eq:iiscatteringbeta}
\beta^2=\frac{e^2}{\epsilon_0 k_BT}\int D_S\left(\varepsilon\right)f\left(1-f\right)d\varepsilon
\end{equation}
where $f$ is the electron distribution and $\epsilon_0$ is the low-frequency dielectric constant. Details on the $\alpha$, $D$ and $B$ parameters are given in the literature \cite{book8}.

At high temperatures, after an inelastic (e.g., electron phonon) scattering event, where the electron scatters from momentum state $k$ to $k^\prime$, the energy of an electron changes, and hence, the electron distribution also changes.  (Note that the distribution may also be perturbed by external forces, such as an electric field or temperature gradient.)  Thus, $f$ becomes a function of $k$, so it must be mapped via the electronic band structure, $\varepsilon \left(k\right)$.  This effect can be shown as a deviation from Fermi-Dirac behavior (Equation \ref{eq:perturbation}).  After some mathematical manipulation, for which details can be found in the literature \cite{book8}, the BTE can be reformulated as:
\begin{equation}
\label{eq:bte_reduced_inelastic_g} g = \frac{S_i \left( g^\prime \right) - \nu \left( \frac{\partial f}{\partial z} \right) - \frac{eF}{\hbar} \frac{\partial f}{\partial k}}{S_o + \nu_{el}} \\
\label{eq:bte_reduced_inelastic_g} 
\end{equation}
\begin{equation}
S_i \left( g^\prime \right) = \int d k^\prime X g \left( k^\prime \right) \left[ s_{inel} \left( k^\prime, k \right) \left[ 1 - f \left( k \right) \right] + s_{inel} \left( k, k^\prime \right) f \left( k \right) \right]
\label{eq:bte_reduced_inelastic_in} 
\end{equation}
\begin{equation}
S_o = \int d k^\prime \left[ s_{inel} \left( k, k^\prime \right) \left[ 1-f \left( k^\prime \right) \right] + s_{inel} \left( k^\prime, k \right) f \left( k^\prime \right) \right] \\
\label{eq:bte_reduced_inelastic_out} 
\end{equation}
Detailed integrated expressions for the scattering in, $S_i$, and scattering out, $S_o$, terms are available in the literature \cite{book8}. The reformulated BTE can then be solved iteratively, using Rode's method \cite{book8,Ager_InN,rode_physrevb_1970,carrier19,fundamental_of_carrier_transport_2009,semiconductor_transport_2000} since $S_i \left( g^\prime \right)$ and $f$ themselves are functions of $g$. First, the Fermi-Dirac distribution can be plugged into the right-hand side of Equation \ref{eq:bte_reduced_inelastic_g} to obtain the first guess, $g_1$, which in turn is used to obtain a new electron distribution to solve for the next guess, $g_2$; this process continues until $g$ converges to a unique value.  Typically, five iterations are required for the perturbation to converge for polar optical phonon scattering in GaAs and InN. More details on Equations \ref{eq:bte_reduced_inelastic_g}-\ref{eq:bte_reduced_inelastic_in} are available in the literature \cite{book8}.

\subsection{Phonon Dispersion}

\begin{figure}
	\subfloat[GaAs]{
		\label{fig:phonon_GaAs}
		\includegraphics[width=0.468\columnwidth]{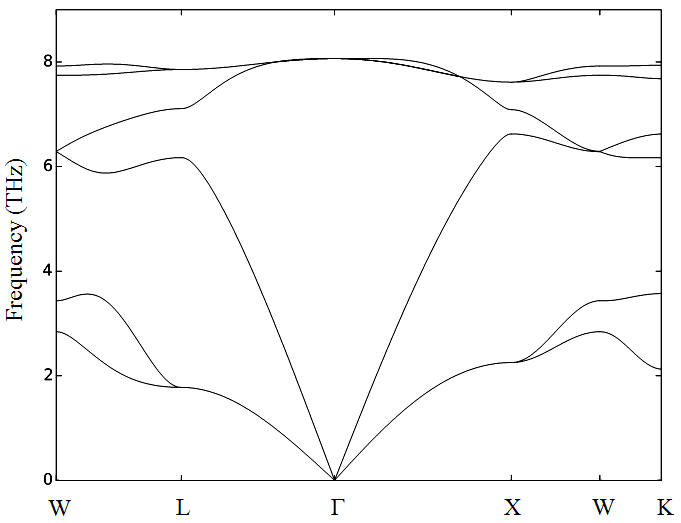}}
	\subfloat[InN]{
		\label{fig:phonon_InN}
		\includegraphics[width=0.475\columnwidth]{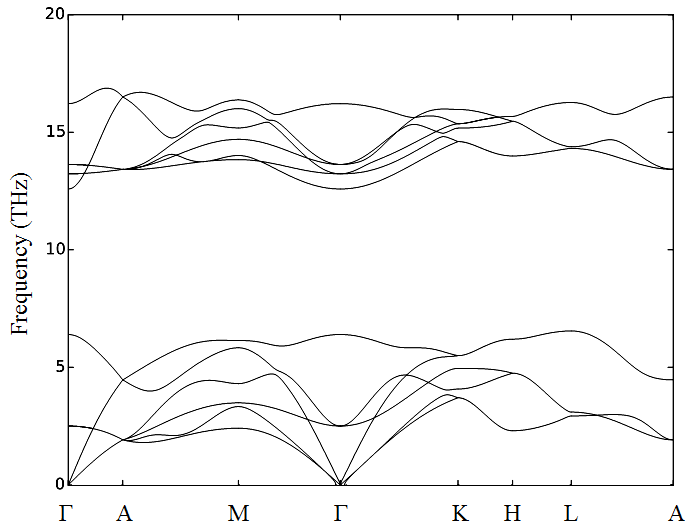}}
	\caption{Phonon band structures of InN and GaAs calculated by Phonopy \cite{phonopy}.}
	\label{fig:phonon}
\end{figure}

Polar optical phonon scattering originates from interactions between electrons and high-frequency optical phonons. They provide the dominant inelastic electron scattering mechanism near (and above) room temperature in compound semiconductors. This is attributed to the high energies of optical phonons being comparable to $k_BT$ at high temperatures. The scattering rates themselves are strongly dependent on the polar optical phonon frequencies. $\omega_{po}$.  These frequencies can be calculated using the Phonopy code \cite{phonopy} which solves for dynamical matrix from the force constants calculated using density functional perturbation theory (DFPT), as implemented in VASP. The phonon band structures for GaAs and InN are shown in Figure \ref{fig:phonon}.

\section{Sensitivity analysis}
\subsection{Sensitivity to the calculated dielectric constants}
\begin{figure}
		\includegraphics[width=0.475\columnwidth]{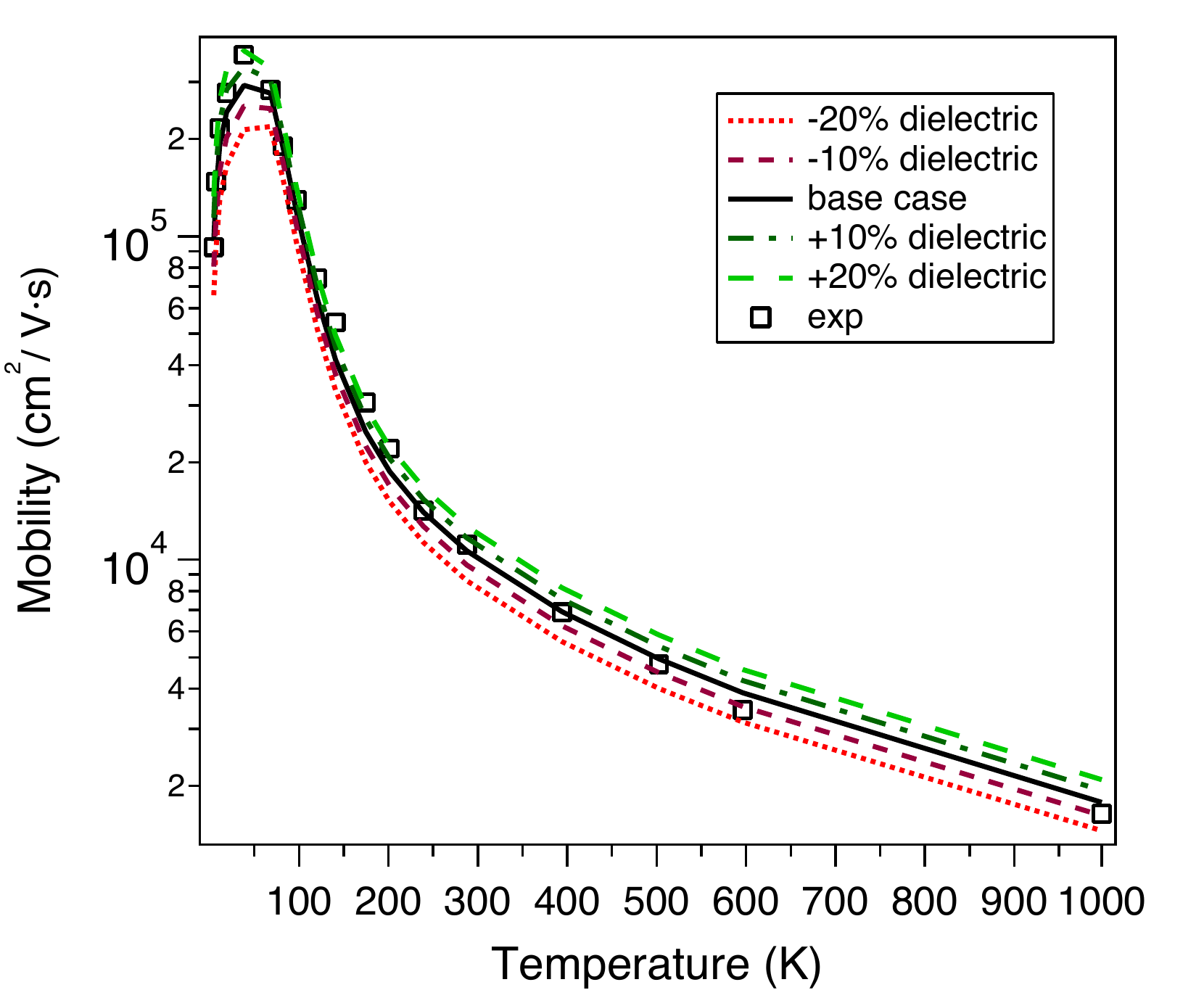}
	\caption{\textcolor{black}{Sensitivity analysis of the mobility of the GaAs pure sample (see Table \ref{tab:GaAs_samples}). We changed here only the static, $\varepsilon_s$, and high frequency, $\varepsilon_\infty$, dielectric constant from -20\% to +20\% from the base values reported in Table \ref{tab:inputs}. The results are sensitive to dielectric constant at low and high temperatures.}}
		\label{fig:GaAs_dielectric_sensitivity}
\end{figure}

\textcolor{black}{We have performed a sensitivity analysis for the calculated mobility of the GaAs pure sample at different dielectric constants. As shown in Figure \ref{fig:GaAs_dielectric_sensitivity}, the result is sensitive to dielectric constants at low and high temperatures but much less sensitive at temperatures in the 100-200 K range. Inaccurate calculation of dielectric constants by -20\% can result in errors of up to -41\% (at 40 K) in the calculated mobility, compared to experimentally measured values; for deviations of +20\% in the dielectric constants, the resulting mobility can increase by up to +43\% (at 5 K). The base values are the ones reported in Table \ref{tab:inputs}, as calculated \emph{ab intitio} and assuming the relaxed structure. This shows the importance of accurate calculation of these constants with at least 5-10\% accuracy.}

\subsection{Sensitivity to the lattice constants}

\begin{figure}
		\includegraphics[width=0.475\columnwidth]{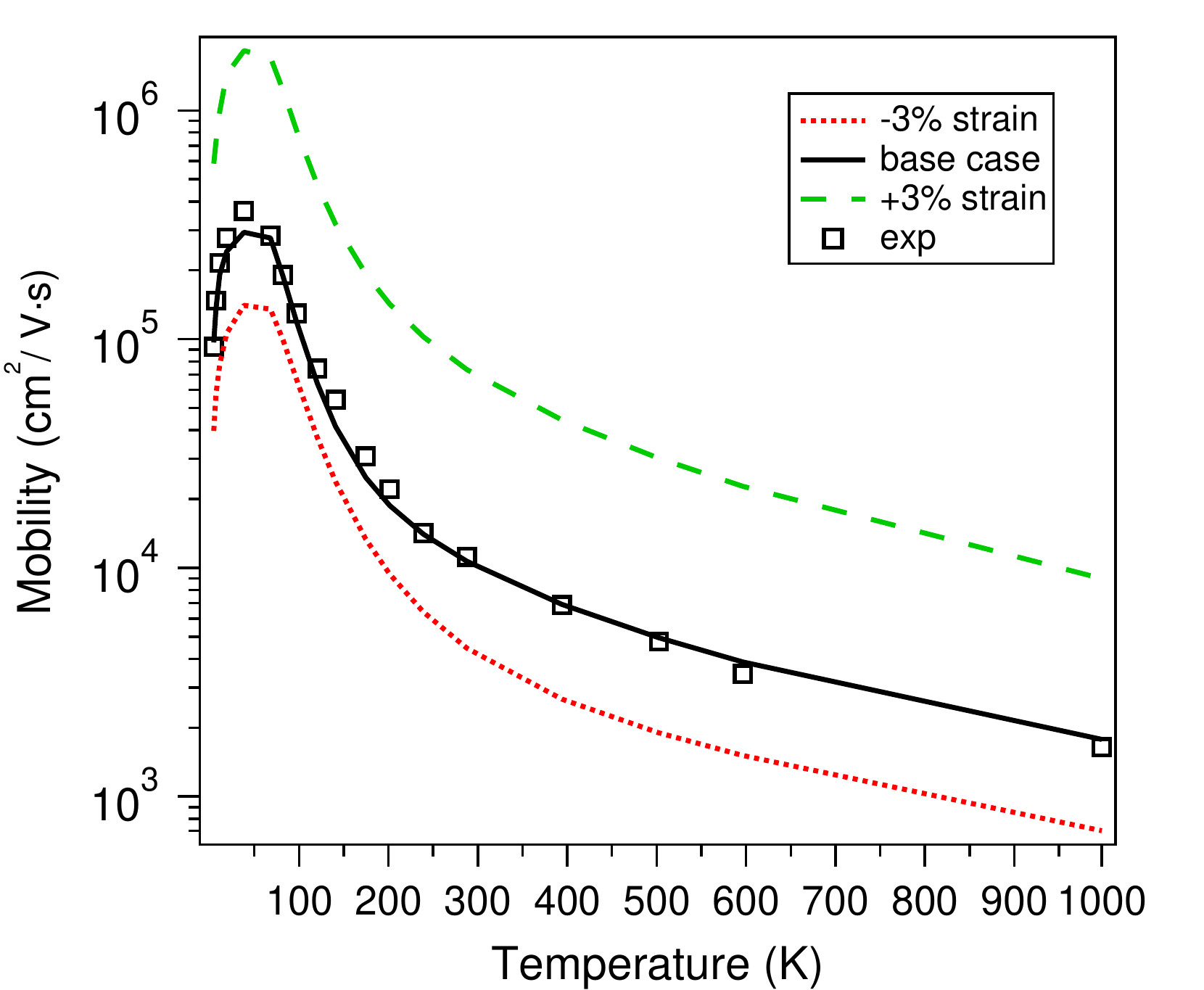}
	\caption{\textcolor{black}{Sensitivity analysis of the mobility of the GaAs pure sample (see Table \ref{tab:GaAs_samples}). We changed here the crystal structure, and subsequently, the newly calculated optical phonon frequencies. The calculated mobility is sensitive to the strain at all temperatures.}}
		\label{fig:lattice_sensitivity}
\end{figure}

\textcolor{black}{We also applied $\pm3\%$ strain to the lattice constant of the relaxed GaAs unit cell, and recalculated the band structure, DOS and optical phonon frequencies to ascertain the effect on the calculated mobility. We assumed that everything else is kept constant according to the base values (see Table \ref{tab:inputs}). According to Figure \ref{fig:lattice_sensitivity}, the calculated mobility is extremely sensitive to the crystal structure. This is mainly due to the impact that the structure has on the band shape (i.e. group velocity of the electrons) since the mobility at any temperature is affected. For example, the GW band structure of -3\% strained GaAs gives an effective mass of 0.026 while that of +3\% strained GaAs gives an effective mass of 0.10. Both of these values are well outside of the range of the reported experimental values (0.064-0.082, see Table \ref{tab:inputs}). Also, these strained structures are extremely unlikely to be relaxed with any functional, since their built-in pressure with GGA-PBE functionals are already 10.66 kB and -74.77 kB, respectively, while the relaxed structure that we  calculated and reported in Table \ref{tab:geometry} has a built-in pressure of only -0.3 kB. Nevertheless, Figure \ref{fig:lattice_sensitivity} shows the importance of accurate calculation of the crystal structure, and subsequently, the band structure (i.e., group velocities).   }

\end{appendix}

\clearpage

\bibliography{faghaninia_2014}


\end{document}